\documentclass[amsmath, amssymb, aps, showpacs, prd, twocolumn, nofootinbib]{revtex4-1}
\usepackage[T1]{fontenc}
\usepackage{ae,aecompl}
\usepackage{enumerate}
\usepackage[colorlinks=true, linkcolor=blue, citecolor=red]{hyperref}
\usepackage{graphicx}
\usepackage{color}
\usepackage[normalem]{ulem}
\usepackage{empheq}
\usepackage[justification=centerlast, singlelinecheck=false, caption=false]{subfig}

\newcommand{\rd}{\mathrm{d}}
\newcommand{\re}{\mathrm{e}}
\newcommand{\Mp}{M_\mathrm{P}}

\newcommand{\mixInd}[3]{{#1}^{#2}_{\hphantom{#2}#3}}
\newcommand*\widefbox[1]{\fbox{#1}}

\renewcommand{\vec}{\boldsymbol}

\begin{document}

\title{Radially stabilized inflating cosmic strings}

\author{Florian Niedermann}
\email[]{florian.niedermann@physik.uni-muenchen.de}
\affiliation{Arnold Sommerfeld Center for Theoretical Physics, Ludwig-Maximilians-Universit\"at, Theresienstra{\ss}e 37, 80333 Munich, Germany}
\affiliation{Excellence Cluster Universe, Boltzmannstra{\ss}e 2, 85748 Garching, Germany\\
~\\}

\author{Robert Schneider}
\email[]{robert.bob.schneider@physik.uni-muenchen.de}
\affiliation{Arnold Sommerfeld Center for Theoretical Physics, Ludwig-Maximilians-Universit\"at, Theresienstra{\ss}e 37, 80333 Munich, Germany}
\affiliation{Excellence Cluster Universe, Boltzmannstra{\ss}e 2, 85748 Garching, Germany\\
~\\}
\date{\today}

\begin{abstract}
In General Relativity, local cosmic strings are well known to produce a static, locally flat spacetime with a wedge removed. If the tension exceeds a critical value, the deficit angle becomes larger than $ 2\pi $, leading to a compact exterior that ends in a conical singularity. 
In this work, we investigate dynamical solutions for cosmic strings with super-critical tensions.
To this end, we model the string as a cylindrical shell of finite and stabilized transverse width and show that there is a marginally super-critical regime in which the stabilization can be achieved by physically reasonable matter.

We show numerically that the static deficit angle solution is unstable for super-critical string tensions. Instead, the geometry starts expanding in axial direction at an asymptotically constant rate, and a horizon is formed in the exterior spacetime, which has the shape of a growing cigar.
We are able to find the analytic form of the attractor solution describing the interior of the cosmic string. In particular, this enables us to analytically derive the relation between the string tension and the axial expansion rate.
Furthermore, we show that the exterior conical singularity can be avoided for dynamical solutions.

Our results might be relevant for theories with two extra dimensions, modeling our universe as a cosmic string with a three-dimensional axis. We derive the corresponding Friedmann equation, relating the on-brane Hubble parameter to the string tension or, equivalently, brane cosmological constant. 
\end{abstract}


\maketitle


\section{Introduction}

Local cosmic strings were first derived as topologically nontrivial solutions of the Abelian Higgs Model by Nielsen and Olesen~\cite{Nielsen:1973cs}. In General Relativity (GR), they give rise to a static geometry which sufficiently far away from the string is locally flat and can be characterized by a deficit angle $\delta$ corresponding to a wedge that has been removed from spacetime. The value of $\delta$ is linearly related to the string tension $ \lambda $ (mass per unit length): $\delta=\lambda / \Mp^2$, with the reduced Planck mass $ \Mp^2 \equiv 1/(8\pi G_\mathrm{N}) $. This spacetime was first studied in \cite{Vilenkin:1981zs, Gott:1985, Hiscock:1985uc}.

Once the tension reaches the critical value $2\pi \Mp^2$, the deficit angle becomes $2 \pi$, thus implying the exterior topology of an infinite cylinder~\cite{Linet:1990fk}. Introducing the dimensionless parameter $ \bar\lambda := \lambda / \left(2\pi\Mp^2\right) $, this critical value corresponds to $ \bar\lambda = 1 $. For even higher values of the string tension, the angular defect exceeds $ 2\pi $; thus, the exterior spacetime of the static solution closes up and ends in a conical singularity\footnote{For both sub- and super-critical tensions, there is a second class of solutions, usually referred to as the ``Melvin'' or ``Kasner'' branch \cite{Laguna:1989rx, Christensen:1999wb}, which has even a curvature singularity in the super-critical case. However, we discard this branch due to its unphysical properties, cf.~Sec.~\ref{sec:static}.}~\cite{Ortiz:1990tn}.
However, the status of these so-called ``super-critical'' or ``super-massive'' solutions remained unclear due to the occurrence of the singular axis away from the string. One way to give a physical meaning to the singularity is to replace it with another (sub-critical) tension string~\cite{Blanco-Pillado:2013axa}.

In this work, we further explore the geometry of super-critical cosmic strings. Instead of introducing additional strings, {\it we relax the assumption of having a static geometry.} A first purely numerical attempt in that direction was made in \cite{Cho:1998xy} by considering a super-critical Nielsen-Olesen (NO) string. There it was found that once the tension $\bar \lambda$ exceeds $\sim 1.6 $, {\it both} the transverse and axial string directions begin to expand at a comparable rate. In this work we will be able to analytically confirm this bound. In contrast to \cite{Cho:1998xy}, we will be mostly interested in describing the remaining parameter space: $1< \bar\lambda \lesssim1.6$. We show that within this ``marginally super-critical'' regime the transverse string dimensions can be stabilized, whereas the axial dimension expands at an asymptotically constant rate.  This fact makes these solutions especially interesting for models with two extra-dimensions according to which the string is promoted to a braneworld describing our universe. Then, the constant axial expansion rate corresponds to a de Sitter on-brane geometry, and having a stabilized transverse dimension is a necessary requirement to obtain a 4D regime in those theories, see e.g.~\cite{Niedermann:2014bqa}.

Furthermore, the analysis of \cite{Cho:1998xy} lacks a detailed discussion of the geometry away from the string. In particular, it was not answered whether the second axis still bears a {\it conical} singularity (which does not lead to asymptotically diverging curvature invariants) as it is the case for the static solution. Our approach allows us to get a complete understanding of both the geometry and the underlying dynamics of the system in the marginally super-critical regime.  For example, we show that the exterior conical singularity can be completely avoided for a dynamical solution and should thus be regarded as an artifact caused by assuming a static geometry. In addition, we show the existence of a horizon between the string and the second (regular) axis. 

In order to technically simplify the problem, we introduce in Sec.~\ref{sec:shell_model} a model which describes the string as a cylindrical shell of fixed circumference $ 2\pi R $. It is argued that this simplified description captures all essential physics as long as we are interested in low energy questions which do not require to resolve the inner structure of the defect. Since the only scale determining the dynamics of the system is the axial expansion rate $ H $, this condition should clearly be satisfied as long as $ HR \ll 1 $.

The shell construction implies the existence of two vacuum regions, one inside and one outside of the cylinder. In Sec.~\ref{sec:coord}, we employ the cylindrical symmetry of this setup to introduce appropriate coordinates, which were originally used by Einstein and Rosen to derive the existence of cylindrical gravitational waves~\cite{EinsteinRosen1937}. Moreover, we derive the junction conditions relating the two vacuum regions across the shell.

As a first consistency check of our description, we reproduce the well-known static deficit angle solution in Sec.~\ref{sec:static}, which --- in the super-critical case --- leads to a second singular axis in the exterior vacuum region. In order to see whether these are stable solutions, we numerically study the time evolution of the system in Sec.~\ref{sec:num_sol}. To that end, we choose initial data close to the static configuration but include cylindrical symmetric gravitational waves in the interior and exterior region close to the shell to provide the system with a non-vanishing initial kinetic energy. We find that the sub-critical string, after emitting the cylindrical waves, settles back to the static deficit angle solution. The super-critical system, however, starts to approach a non-static solution instead. {\it This result proves that the static super-critical solution is not stable under perturbations.}. Moreover, the numerical results allow to infer several properties of the new dynamical attractor solution in the marginally super-critical regime defined by $1< \bar\lambda \lesssim1.6$:
\begin{itemize}
	\item The geometry expands in axial direction at a constant rate. 
	
	\item There is a horizon in the exterior region.
	
	\item The exterior space is cigar-shaped and expands, whereas the interior space is nearly flat and can be stabilized.
\end{itemize}
Let us emphasize that for $ \bar \lambda $ close to 1, the observed expansion rate satisfies $ HR \ll 1 $. Therefore, in that regime we expect all the results to be completely insensitive to the microscopic inner structure of the string. They should thus equally hold for other regularizations like a full cylinder for which the tension would be smeared out over the interior region, or the original UV model described in~\cite{Nielsen:1973cs}.

For even larger values of the tension, $ \bar\lambda \gtrsim1.6 $, we find that the azimuthal pressure required to stabilize the shell's circumference violates the Null Energy Condition (NEC). Hence, in this regime a stabilization cannot be achieved by physical degrees of freedom. Since at this point $ HR$ is already close to one, it is a priori not clear whether this result would still hold for a more realistic NO-like string. However, it turns out that our NEC-bound is in nice agreement with the one derived in~\cite{Cho:1998xy}, where the full radial profile was resolved. This is a strong indication that our simplified model can in fact be successfully used in the {\it whole} stabilizable (i.e. marginally super-critical) regime to capture the essential physics --- instead of the more complicated microscopic NO-system. This result is also in agreement with the idea of ``topological inflation''~\cite{Vilenkin:1994pv, Linde:1996rb, Linde:1994wt}. In that context, it is argued that once $HR \sim 1$, the interior space of the defect starts to inflate in {\it both} axial and radial direction at the same rate, or in other words, there is a de Sitter phase inside the cosmic string. This is plausible because at this point its boundary lies outside the corresponding horizon and thus the interior is causally disconnected from the exterior, which makes it locally equivalent to a pure de Sitter universe.

We derive the analytic form of the attractor solution in the interior of the shell in Sec.~\ref{sec:analytic} by making an appropriate scaling ansatz for the metric. This in turn enables us to derive the relation between the string tension $\lambda$ and $H$ analytically. In the marginally super-critical regime, we find up to small corrections of order $(HR)^2$:
\begin{equation}\label{eq:main_result}
HR \approx \bar\lambda - 1 \;.
\end{equation}
The analytic result for the interior geometry can be mapped by a coordinate transformation, described in Appendix~\ref{ap:cigar_coord}, to a solution discussed earlier by Witten~\cite{Witten:1982} and Gregory~\cite{Gregory:2003xf}. To our knowledge, this is the first time that this solution has been matched to a specific matter model.

Furthermore, we are able to show that the conical singularity in the exterior is an artifact caused by assuming a static geometry. More precisely, in Sec.~\ref{sec:remove_conical_sing} we demonstrate that the singularity can be completely avoided by choosing the initial conditions appropriately. In that case the exterior space ends radially in a smooth and elementary flat axis. The existence of a singularity-free inflating solution was already anticipated in~\cite{Vilenkin:1994pv, Kaloper:2007ap}. Moreover, we argue that the value of $H$ is completely independent of the choice of initial conditions and solely depends on the string parameters $R$ and $\lambda$, as expected for an attractor solution. A corresponding parameter plot, summarizing our results, is discussed in Sec.~\ref{sec:parameter_plot}.

For a cosmic string formed during a phase transition in the early universe, say at the GUT scale, we would generically expect a sub-critical tension of order $\bar \lambda \sim 10^{-6}$. However, in~\cite{Ortiz:1990tn} it was argued that super-critical cosmic strings could also arise at this scale when the coupling between scalar and gauge fields is very weak (in a NO-framework). In order to further clarify their phenomenological status for standard cosmology, we review arguments previously given by Thorne \cite{PhysRevD.46.2435}. In this context, it is shown in Appendix~\ref{ap:ext_geom} that an open cylindrical geometry cannot evolve classically into a closed one. This result implies that the formation of super-critical strings cannot be described within General Relativity. So far it is not clear whether their formation through a quantum-mechanical tunneling process would lead to phenomenologically relevant effects. 

However, it is worth pointing out that these solutions might be interesting in the context of higher dimensional models, where the string is promoted to a 3-brane representing our universe. A corresponding generalization of equation~\eqref{eq:main_result} can be understood as a modified Friedmann equation. The axial expansion rate $H$ then plays the role of the ordinary Hubble parameter describing the spatial expansion of our universe. In that context, the regime $HR \ll 1$, corresponding to a marginally super-critical tension, is the most interesting one as it is enforced by a huge separation between the cosmological length scale $H^{-1}$ and the microscopic scale $R$ given by the thickness of the string. The six dimensional setup in the case of a pure tension brane is discussed in Section~\ref{sec:6D}. We draw our conclusions in Section \ref{sec:conclusion}.

Finally, in Appendix \ref{ap:EFT} we show that the super-critical cosmic strings covered by our analysis can indeed be described consistently within classical GR, as long as the transverse string size $ R $ is much larger than the Planck length, since this ensures a sub-Planckian 4D energy density.

\section{Cylindrical shell model}
\label{sec:shell_model}

Strings with a sub-critical tension can consistently be modeled as infinitely thin objects, with the position of the string corresponding to a conical singularity. In the super-critical case of interest, however, the deficit angle exceeds $ 2\pi $ and so one can no longer treat the string as infinitely thin because in this case the space-time would simply disappear. Hence, the question whether the static wedge-geometry is a stable solution for super-critical tensions can only be investigated in the more realistic situation of a regularized string, i.e.\ one that has a finite width.
Physically, this non-zero thickness can also be motivated if one imagines the string to be dynamically formed as a vortex solution in a NO-like model~\cite{Nielsen:1973cs}. However, for simplicity, we will not consider such a UV-complete construction, but instead model the regularized string as a cylindrical shell of finite circumference $ 2\pi R $ and tension $ \lambda $. This regularization is widely used for codimension-two objects, e.g.\ \cite{Peloso:2006cq,Kaloper:2007ap}.

Furthermore, we will assume the thickness of the string to be constant, so it is only allowed to expand (or collapse) in axial direction. Technically, this is achieved by choosing an appropriate azimuthal pressure\footnote{Actually, this only fixes the circumference of the shell, but we will see a posteriori that it also sufficiently stabilizes the whole interior cross-sectional area.} $ p_\phi $. Physically, it means that we are effectively working well below the energy scale at which the string was formed in an actual UV model and at which its inner structure could be probed. We will confirm a posteriori that this stabilization could correspond to a healthy microscopic theory by showing that the required source fulfills the standard energy conditions.

Even though this shell construction corresponds to quite a different radial profile than a typical vortex solution would give, we expect the low energy questions that we are addressing to be insensitive to these details. 
To be more precise, as long as the axial expansion rate $H$ is much smaller than the inverse transverse length scale $R$,
\begin{align}\label{eq:consistent}
	HR \ll 1\;,
\end{align}
we do not need to resolve the inner structure of the string to make accurate predictions. We will see a posteriori that \eqref{eq:consistent} is fulfilled for our solutions if the tension is close to its critical value.

Another justification for that regularization comes from the fact that in the static case, which will be reviewed in Sec.~\ref{sec:static}, this simplified model correctly reproduces all the essential geometrical features of the complete NO analysis~\cite{Laguna:1989rx}. Moreover, in the dynamical case we will find remarkable agreement with a specific quantitative result derived in the NO-setup in~\cite{Cho:1998xy}, namely the stability bound on $\bar \lambda$ for the transverse string directions.

\section{Coordinates and geometry}
\label{sec:coord}

Since we are considering an infinitely long straight cylindrical shell, the system has whole cylinder symmetry, i.e.\ rotational symmetry about the cylinder axis, as well as translational symmetry along the axis.
As discussed in~\cite{Thorne:1965} (also see \citep[chap. 22]{Stephani}), in this case the general metric can always be brought into the form
\begin{equation} \label{eq:met_general}
	\rd s^2 = \re^{2(\eta^* - \alpha^*)} \left( -\rd t^{*2} + \rd r^{*2} \right) + \re^{2\alpha^*} \rd z^2 + \re^{-2\alpha^*} W^2 \rd\phi^2 \,.
\end{equation}
Here, $ \phi \in [ 0, 2\pi ) $ and $ z \in (-\infty, \infty) $ are coordinates in angular and axial direction, respectively, and the functions $ \alpha^*, \eta^*, W $ only depend on the temporal and radial coordinates $ (t^*, r^*) $. One virtue of this ansatz is that radial light-rays correspond to $ \rd  t^* = \pm \rd r^* $, thus making the causal structure evident in $ t^* $-$ r^* $-diagrams, a fact which we will often use in our analysis.

The form of the metric \eqref{eq:met_general} is invariant under the residual coordinate transformations
\begin{subequations} \label{eq:coord_trafo}
\begin{align}
	(t^*,r^*) & \mapsto (t, r) \,,\\
	\eta^* & \mapsto \eta = \eta^* - \frac{1}{2} \ln \left[ (\partial_{r^*} r)^2 - (\partial_{t^*} r)^2 \right] \,, \label{eq:coord_trafo_eta}
\end{align}
\end{subequations}
where $ t $ and $ r $ are any functions of $ (t^*, r^*) $ satisfying\footnote{For the Jacobian not to vanish, they should also obey $ (\partial_{t^*}r)^2 - (\partial_{r^*}r)^2 \neq 0$.}
\begin{align} \label{eq:cond_trafo}
	\begin{pmatrix} \partial_{t^*} r \\ \partial_{r^*} r \end{pmatrix}
	= \pm
	\begin{pmatrix} \partial_{r^*} t \\\partial_{t^*} t \end{pmatrix}\,.
\end{align}
%
This implies an integrability condition for $ r $:
\begin{equation}
\label{eq:cond_trafo_2}
	- \partial_{t^*}^2 r + \partial_{r^*}^2 r =  0\,.
\end{equation}
In vacuum, Einstein's equations imply
\begin{equation}
\label{eq:1D_wave}
	-\partial_{t^*}^2 W + \partial_{r^*}^2 W  = 0 \,,
\end{equation}
which allows to fix the remaining gauge freedom by choosing\footnote{By using $ r $ as a spatial coordinate, we implicitly assume that the gradient of $ W $ is space-like. We will come back to this important subtlety in Section \ref{sec:supercrit}.}
\begin{equation}
\label{eq:ER_gauge}
	r = W(t^*, r^*) \,.
\end{equation}
In general, however, the right hand side of \eqref{eq:1D_wave} receives a contribution $ \propto T^{t}_{\hphantom{t}t} + T^r_{\hphantom{r}r} $. Thus, while the gauge \eqref{eq:ER_gauge} can be used inside and outside the cylinder separately, the condition \eqref{eq:cond_trafo_2} would then not be fulfilled across its surface. Consequently, the interior and exterior coordinate patches are not continuously connected. To distinguish them, we put tildes on the interior coordinates and metric functions. The interior and exterior metric then read
\begin{subequations} \label{eq:met_ER}
\begin{align}
	\rd \tilde s^2 & = \re^{2(\tilde \eta - \tilde \alpha)} \left( -\rd \tilde t^2 + \rd \tilde r^2 \right) + \re^{2\tilde \alpha} \rd z^2 + \re^{-2\tilde \alpha} \tilde r^2 \rd\phi^2 \,, \label{eq:met_ER_in}\\
	\rd s^2 & = \re^{2(\eta - \alpha)} \left( -\rd t^2 + \rd r^2 \right) + \re^{2\alpha} \rd z^2 + \re^{-2\alpha} r^2 \rd\phi^2 \,,
\end{align}
\end{subequations}
respectively. This is the ansatz that was used by Einstein and Rosen to derive the exact vacuum solutions describing cylindrically symmetric waves in GR \cite{EinsteinRosen1937}. The vacuum Einstein equations in these ``Einstein Rosen coordinates'' become
\begin{subequations}\label{eq:ER_vac}
\begin{align}
	\partial_t^2\alpha &= \partial_r^2\alpha + \frac{1}{r}\partial_r\alpha \,, \label{eq:ER_vac_alpha} \\
	\partial_r \eta &= r \left[ \left(\partial_t\alpha\right)^2 + \left(\partial_r\alpha\right)^2 \right] \,, \label{eq:ER_vac_eta_r} \\
	\partial_t \eta &= 2r \left(\partial_t\alpha \right) \left(\partial_r\alpha\right) \,, \label{eq:ER_vac_eta_t}
\end{align}
\end{subequations}
and similarly for the interior. The remarkable fact that $ \alpha $ obeys the linear cylindrical wave equation makes this coordinate choice unique and especially convenient for numerical implementation.

The symmetry axis is located in the interior coordinate patch at $ \tilde r = 0 $.
Regularity at this axis and elementary flatness, i.e.\ absence of a conical singularity, requires
\begin{subequations}\label{eq:reg_axis}
\begin{align}
	\lim_{\tilde r \to 0} \partial_{\tilde r} \tilde \alpha &= 0 \qquad \text{and} \label{eq:reg_axis_alpha}\\
	\lim_{\tilde r \to 0} \tilde \eta &= 0\;, \label{eq:reg_axis_eta}
\end{align}
\end{subequations}
respectively.

\subsection{Induced geometry} \label{sec:ind_metric}

The induced metric on the cylindrical shell is
\begin{equation}\label{eq:induced_met}
	\rd s^2_\mathrm{(ind)} = -\rd \tau^2 + \re^{2\alpha_0} \rd  z^2 + R^2 \rd\phi^2 \,,
\end{equation}
where here and henceforth the subscript ``0'' denotes evaluation at the position of the shell. The proper time $ \tau $ on the surface is related to the interior and exterior time coordinates via
\begin{equation}\label{eq:timeRelation}
	\rd \tau = \frac{\re^{-\alpha_0}}{\gamma} \rd t = \frac{\re^{-\tilde\alpha_0}}{\tilde\gamma} \rd \tilde t \,,
\end{equation}
where
\begin{equation}\label{eq:defGamma}
	\gamma := \frac{\re^{-\eta_0}}{\sqrt{1-\dot r_0^2}}
	\qquad\text{and}\qquad
	\tilde \gamma := \frac{\re^{-\tilde \eta_0}}{\sqrt{1-\dot {\tilde r}_0^2}}\,.
\end{equation}
The functions $\tilde r_0(\tilde t)$ and $r_0(t)$ describe the radial position of the shell in the two coordinate patches, and $\dot r_0 := \rd r_0 /\rd t$, $\dot {\tilde{r}}_0 := \rd \tilde r_0 /\rd \tilde t$.

In order to have a well defined regularization of the infinitely thin cosmic string, we assume the proper circumference of the cylinder to be stabilized:
\begin{equation}\label{eq:defR}
	R := r_0 \re^{-\alpha_0} = \tilde r_0 \re^{-\tilde \alpha_0} = \text{const.}
\end{equation}
As already mentioned, on a fundamental level this would be enforced by some underlying UV physics that gave rise to the cosmic string.
Effectively, working well below this UV scale at which the inner structure of the string could be probed, it can be achieved by assuming a suitable azimuthal pressure component $ p_\phi $. We will check a posteriori whether this pressure is physically reasonable, i.e.\ whether it satisfies the Null Energy Condition.

The surface energy momentum tensor on the shell is given by
\begin{equation} \label{eq:Tmn}
	\mixInd{T}{m}{n} = \frac{1}{2\pi R} \text{diag}\left( -\lambda, -\lambda, p_\phi \right) \,,
\end{equation}
where the overall factor ensures that $ \lambda $ is the one-dimensional string tension.
Throughout this work we will assume that $ \lambda \geq 0 $.
Let us, for later convenience, also introduce the dimensionless quantities
\begin{align}
	\bar\lambda := \frac{\lambda}{2\pi\Mp^2} \,, && \bar p_\phi := \frac{p_\phi}{2\pi\Mp^2} \,.
\end{align}
Fixing $ R $ implies that the 3D energy conservation equation for this source simply becomes $ \lambda = \text{constant} $. The pressure $ p_\phi $ will in general be time-dependent, and its value will be inferred from one of the junction conditions, see below. Furthermore, the entire dynamics of the induced metric \eqref{eq:induced_met} is now encoded in the single function $ H := \rd \alpha_0 / \rd \tau = \rd \tilde \alpha_0 / \rd \tau $, measuring the expansion rate of the string in axial direction.

For future reference, note that the stabilization condition implies
\begin{align} \label{eq:r0Dot_rel}
	\dot r_0 = \frac{H R}{\gamma} \,, && \dot{\tilde r}_0 = \frac{H R}{\tilde\gamma} \,, 
\end{align}
which allows us to rewrite \eqref{eq:defGamma} as
\begin{align} \label{eq:gamma}
	\gamma = \sqrt{\re^{-2\eta_0} + H^2 R^2} \, , && \tilde\gamma = \sqrt{\re^{-2\tilde\eta_0} + H^2 R^2} \, .
\end{align}
%

\subsection{Super-criticality} \label{sec:supercrit}

The interpretation of $ r $ and $ \tilde r $ as radial coordinates implicitly assumes that the gradient $ \nabla W := (\partial_{t^*}W, \partial_{r^*}W) $ of the function $ W $ in the original metric \eqref{eq:met_general} was space-like. If it had been time-like, the coordinate $ r $ defined by~\eqref{eq:ER_gauge} would in fact be a temporal coordinate. This agrees with the observation that, according to \eqref{eq:coord_trafo_eta}, $ \exp(2\eta^*) $ would then have changed sign under the coordinate transformation, and $ t $ and $ r $ would thus have interchanged their temporal and spatial character in \eqref{eq:met_ER}.

In order to have a smooth symmetry axis in the interior, and hence a well defined regularization of the cosmic string, we have to assume that $ \nabla W $ was space-like in the interior region. Furthermore, since $ \tilde r $ should take positive values, $ \nabla W $ had to be outward pointing. As discussed in more detail in Appendix~\ref{ap:ext_geom}, the character of $ \nabla W $ in the exterior is then fixed by the amount of string tension $ \lambda $ that is localized on the cylindrical shell. There are three cases:

\begin{enumerate}[(i)]
\item 
For $ \lambda $ small enough, viz.\
\begin{equation} \label{eq:sub_crit}
	\bar\lambda < \tilde\gamma - \left | H \right | R \,, 
\end{equation}
the exterior gradient $ \nabla W^\mathrm{(ext)} $ is also space-like and outward pointing, leading to a conical, but infinite exterior geometry with $ r \in (r_0, \infty) $.

\item
In the intermediate regime
\begin{equation}\label{eq:crit}
	\tilde\gamma - \left | H \right | R \leq \bar\lambda \leq \tilde\gamma + \left | H \right | R \,, 
\end{equation}
$ \nabla W^\mathrm{(ext)} $ is time-like\footnote{Or light-like, if one of the bounds is saturated. In that case the coordinate transformation would be singular.} and $ r $ is thus a temporal coordinate. We exclude this ``critical'' case from our current analysis.

\item
If the tension is large enough,
\begin{equation} \label{eq:super_crit}
	\bar\lambda > \tilde\gamma + \left | H \right | R \,, 
\end{equation}
then $ \nabla W^\mathrm{(ext)} $ is again space-like but \textit{inward} pointing. Thus, $ r $ is again a spatial coordinate; but now it decreases as one moves away from the cylinder surface. In principle, there could be some $ r_\mathrm{min} > 0 $, at which $ r $ starts increasing again. However, this would imply that $ \nabla W $ changed character from inward to outward pointing at $ r_\mathrm{min} $, and one can show that this is not possible in vacuum, see Appendix \ref{ap:ext_geom}. Thus, $ r $ has the finite range $ r \in (r_0, 0) $ and at the point $ r=0 $ there will be a second axis, which can generically be singular. 
\end{enumerate}

We will refer to the first and third case as ``sub-'' and ``super-critical'', respectively. In the static case $ H \to 0 $ and $ \tilde\gamma \to 1 $, and so the conditions take the form (i) $ \bar\lambda < 1 $ and (iii) $ \bar\lambda > 1 $, while the critical range (ii) degenerates to $ \bar{\lambda} = 1 $, cf. Sec.~\ref{sec:static}. In the present work, we are mainly interested in the super-critical regime (iii).


\subsection{Junction conditions} \label{sec:junc_cond}

The vacuum Einstein field equations \eqref{eq:ER_vac} have to be supplemented by Israel's junction conditions \cite{Israel:1966, Israel:1967}, linking the interior and exterior geometries across the cylinder surface:
\begin{equation} \label{eq:israel}
	\mixInd{T}{m}{n} = \Mp^2 \left( [\mixInd{K}{p}{p}] \mixInd{\delta}{m}{n} - [ \mixInd{K}{m}{n} ] \right) \,.
\end{equation}
Here, $[X] := X - \tilde{X}$, and $K_{mn}$ is the (pullback of the) extrinsic curvature tensor.
The outward-pointing normal vectors in the interior and exterior are given by
\begin{align}\label{eq:normal_vec}
	\tilde{n}^\mu = \tilde{\gamma} \re^{\tilde{\alpha}_0} \left( \dot{\tilde{r}}_0, 1, 0,0 \right) && \text{and} &&
	n^\mu = \sigma \gamma \re^{\alpha_0} \left( \dot{r}_0 , 1, 0,0 \right) \,,
\end{align}
respectively, with $ \sigma = \pm 1 $. In order for the normal vector $ n^\mu $ to be outward-pointing, $ \sigma $ has to be $ +1 $ in the sub-critical case. But for super-critical tensions, the exterior radial coordinate decreases as one moves away from the cylinder, so in that case one has to choose $ \sigma = -1 $.

Using these normal vectors, it is straightforward to show that the non-vanishing components of $ \mixInd{K}{m}{n} $ are\footnote{In these formulas, evaluation at the surface should of course be performed \textit{after} taking all occurring $ r $-derivatives.}
\begin{subequations}
	\begin{align}
		\mixInd{K}{0}{0} &= \frac{\sigma\gamma}{R} \frac{r_0 \ddot{r}_0}{1 - \dot{r}_0^2}  + n^\mu \partial_\mu \left(\eta_0 - \alpha_0 \right) \,,\\
		\mixInd{K}{i}{j} &= n^\mu \partial_\mu\alpha_0 \, \mixInd{\delta}{i}{j} \,,\label{eq:extrinsic_ij}\\
		\mixInd{K}{\phi}{\phi} &= \frac{\sigma\gamma}{R} - n^\mu\partial_\mu\alpha_0 \,.
		\label{Kcompsout}
	\end{align}
\end{subequations}
The components of $ \mixInd{\tilde K}{m}{n}$ have the same form, but with tildes on all quantities and $ \sigma \to +1 $.
Plugging this and~\eqref{eq:Tmn} into \eqref{eq:israel}, the $ (0, 0) $-component of the junction conditions becomes\footnote{Note that without choosing the correct sign $ \sigma = -1 $ in the super-critical case, this equation would imply $ \bar\lambda < \tilde\gamma $, in contradiction to the condition \eqref{eq:super_crit}.}
\begin{equation} \label{eq:00_junc_cond}
	\bar\lambda = \tilde\gamma - \sigma\gamma \,.
\end{equation}
The $ (i, j) $-components, after eliminating $ \bar\lambda $ using \eqref{eq:00_junc_cond} as well as $ \eta $ by means of the vacuum equations \eqref{eq:ER_vac} in the limit $ r \to r_0 $, and expressing everything in terms of the intrinsic cylinder quantities $ H $ and $ R $, can be written as
\begin{equation} \label{eq:ij_junc_cond}
	\frac{\rd H}{\rd \tau} R^2 = \left[ \sigma \gamma f(\chi, \xi) - \tilde{\gamma}f(\tilde\chi, \tilde\xi) \right ] \left( \frac{\sigma}{\gamma} - \frac{1}{\tilde\gamma} \right)^{-1}\,,
\end{equation}
where
\begin{subequations}
\begin{align}
	f(\chi, \xi) & := 1 - \chi - (1 - \xi )^2 \left( 1 - \chi \right )^2 \,, \\
	\xi & := r_0 \partial_r\alpha_0\,, \qquad \chi := \left (\frac{HR}{\gamma}\right )^2 \,.
\end{align}
\end{subequations}

The complete set of equations of motion consists of the vacuum field equations \eqref{eq:ER_vac} in the interior and exterior region, the dynamical (second order in time) junction condition \eqref{eq:ij_junc_cond}, and energy conservation ($ \lambda = \text{const.} $), supplemented by the boundary conditions \eqref{eq:reg_axis} (as well as appropriate boundary conditions for the exterior domain which will be discussed later). Equation \eqref{eq:00_junc_cond} is a constraint, i.e.\ it only contains first time derivatives, and only has to be imposed at the initial moment of time. Its conservation is guaranteed by the Gauss-Codazzi and vacuum field equations \cite{Israel:1966, Israel:1967}, and will serve as an important consistency check for the numerical implementation.

Finally, the $ (\phi, \phi) $-junction condition determines the azimuthal pressure $ p_\phi $ that is needed to keep the circumference of the cylinder constant. A similar calculation as before yields
\begin{equation} \label{eq:phiphi_junc_cond}
	\bar p_\phi = \sigma \gamma g(\chi, \xi) - \tilde{\gamma} g(\tilde\chi, \tilde\xi) \,,
\end{equation}
with
\begin{equation}
	g(\chi, \xi) := 2 \left ( \chi - \chi\xi + \xi \right ) \,.
\end{equation}

\section{Static solution} \label{sec:static}

Before investigating dynamical solutions, let us first briefly review the much simpler and well-known case of static cosmic string geometries \cite{Vilenkin:1981zs, Gott:1985, Hiscock:1985uc}. After setting to zero all time-derivatives, the vacuum equations \eqref{eq:ER_vac} can easily be integrated, yielding
\begin{align}
	\alpha(r) = \xi \ln\left(\frac{r}{r_0}\right) \,, && \eta(r) = \xi^2 \ln\left(\frac{r}{r_0}\right) + \eta_0 \,,
\end{align}
where we already used a local rescaling of $ t $ and $ r $ to set $ \alpha_0 = 0 $. The same holds in the interior, but here the regularity conditions \eqref{eq:reg_axis} imply $ \tilde \xi = \tilde \eta_0 = 0 $, so the geometry inside the cylinder is Minkowski.
The two constants $ \xi, \eta_0 $ and the azimuthal pressure $ \bar p_\phi $ are then determined by the junction conditions \eqref{eq:00_junc_cond}, \eqref{eq:ij_junc_cond} and \eqref{eq:phiphi_junc_cond}\footnote{Equation \eqref{eq:ij_junc_cond} is a quadratic equation in $ \xi $, which has the second solution $ \xi = 2 $, usually referred to as the ``Melvin'' or ``Kasner'' branch \cite{Laguna:1989rx, Christensen:1999wb}. However, \eqref{eq:phiphi_junc_cond} would then imply $ p_\phi \neq 0 $ for $ \lambda = 0 $, which is why we discard this branch.}:
\begin{align} \label{eq:static_constants}
	\eta_0 = -\ln \left| 1 - \bar\lambda \right| \,, && \xi = 0 = \bar p_\phi \,.
\end{align}
Hence, the metric functions in the exterior are also constant, and so the spacetime around the string is locally flat as well. However, the nonzero value of $ \eta_0 $ corresponds to a nontrivial global geometrical effect. This can be seen explicitly after rescaling coordinates according to $ (t^*, r^*) = (\re^{\eta_0}t, \sigma\re^{\eta_0}(r - r_0) + r_0) $. Here, the sign was chosen such that the new radial coordinate $ r^* $ is again \textit{increasing} for super-critical tensions as well, and the shift makes the metric continuous across the shell. Hence, the spacetime is again covered by a single coordinate patch, in which the metric reads:
\begin{equation}
	\rd s^2 = -\rd t^{*2} + \rd r^{*2} + \rd z^2 + W(r^*)^2 \rd \phi^2 \,,
\end{equation}
with
\begin{equation}
W(r^*) = 
\begin{cases}
r^* & \mathrm{for} \quad r^* \leq r_0\,,\\
\left( 1 - \bar{\lambda} \right) r^* + \bar{\lambda} r_0 & \mathrm{for} \quad r^* > r_0 \,.
\end{cases}
\end{equation}
While the ratio of physical circumference to radius equals $ 2\pi $ inside, it is smaller outside, corresponding to a conical geometry with defect angle $ 2\pi \bar{\lambda} \equiv \lambda / \Mp^2 $.

For super-critical tensions $ \bar\lambda > 1 $, the physical circumference decreases as one moves away from the string and vanishes for some $ r_1 > r_0 $. This means that there is a second axis at this point in the exterior, and because the regularity condition \eqref{eq:reg_axis_eta} is violated\footnote{Unless $ \bar\lambda = 2 $; we discard this exceptional case in our discussion.}, there is a conical singularity at $ r_1 $.

Even though these static solutions do exist in the super-critical case, it is not a priori clear whether they are stable (attractor) solutions. In order to answer this question, we next investigate general time-dependent solutions.

\section{Numerical results} \label{sec:num_sol}

Although the vacuum equation \eqref{eq:ER_vac_alpha} is linear, the complete system is still highly non-linear due to the junction conditions \eqref{eq:00_junc_cond}, \eqref{eq:ij_junc_cond} and equations \eqref{eq:ER_vac_eta_r}, \eqref{eq:ER_vac_eta_t}, which determine $ \tilde\eta, \eta $ --- and thus $ \tilde\gamma, \gamma $. We therefore solve these equations numerically. 

The numerical solutions are obtained by specifying appropriate initial data, as discussed below, and integrating forward in time using the dynamical equations \eqref{eq:ER_vac_alpha}, \eqref{eq:ij_junc_cond} for $ \alpha $ (and $ \tilde\alpha $), as well as the constraints~\eqref{eq:ER_vac_eta_r}, \eqref{eq:ER_vac_eta_t} in the limit $ r \to r_0 $ ($ \tilde r \to \tilde r_0 $) for $ \eta_0 $ ($ \tilde \eta_0 $)\footnote{The radial profiles of $ \eta, \tilde\eta $ can be obtained as well using \eqref{eq:ER_vac_eta_r} or \eqref{eq:ER_vac_eta_t} in the bulk; but they decouple from the rest and are thus not needed --- only $ \eta_0, \tilde\eta_0 $ enter via the junction conditions.}. The constraint~\eqref{eq:00_junc_cond} is only enforced at initial time. Analytically, it is conserved automatically\footnote{Note that we also use energy conservation, which takes the form $ \lambda = \text{constant} $.} and can thus be used as an important consistency check on the numerics at later times.

The discretization scheme and numerical algorithm is the same as in~\cite{Niedermann:2014bqa}, and is explained in detail there in Appendix C.
In what follows, we first discuss the choice of initial data and boundary conditions and then present the results.

\subsection{Initial data and boundary conditions} \label{sec:init_cond}

The numerical integration starts at some initial time, which --- without loss of generality --- we choose to be $ \tilde t_i = t_i = \tau_i = 0 $. We will denote all functions evaluated at this time with a subscript $ i $.

First of all, we must specify the initial radial profiles $ \tilde\alpha_i(\tilde r) $ and $ \alpha_i(r)$.
Since our first main objective is to check whether the static solutions discussed in Sec.~\ref{sec:static} are stable, we choose the corresponding flat profile as initial data:
\begin{equation}
\label{eq:initProfileAlpha}
	\tilde\alpha_i(\tilde r) = 0 = \alpha_{i}(r) \, .
\end{equation}
Consequently, the initial radial coordinate position of the cylinder is
\begin{equation}
	\tilde r_{0i} = r_{0i} = R \,.
\end{equation}

Next, we have to choose initial velocity profiles $ \partial_{\tilde t}\tilde\alpha_i(\tilde r) $ and $ \partial_t\alpha_i(r) $. If we set these to zero as well, the solution will remain static for all times. This is of course not what we are interested in, so we will choose some non-zero profile functions.

In the interior, regularity at the axis \eqref{eq:reg_axis_alpha} implies
\begin{equation}
	\partial_{\tilde r}\partial_{\tilde t}\tilde\alpha_i(0) = 0.
\end{equation}
At the cylinder surface, the velocity profile is related to the initial expansion rate $ H_i $ via
\begin{equation}
\label{eq:alpha0TildeDot_init}
	\partial_{\tilde t} \tilde \alpha_{0i} = \frac{\rd \tilde \alpha_{0i} }{\rd \tilde t} = \frac{H_i}{\tilde\gamma_i}
\end{equation}
where the first equality uses $ \partial_{\tilde r}\tilde\alpha_i = 0 $ which is satisfied for our choice \eqref{eq:initProfileAlpha}. The most general initial velocity can thus be written as
\begin{equation}
	\partial_{\tilde t} \tilde \alpha_i(\tilde r) = \frac{H_i}{\tilde\gamma_i} \, \tilde F\left(\frac{\tilde r}{R}\right) \,,
\end{equation}
and similarly
\begin{equation}
\label{eq:initProfileAlphaDotOut}
	\partial_t \alpha_i(r) = \frac{H_i}{\gamma_i} \, F\left(\frac{r}{R}\right) \,,
\end{equation}
with some profile functions $ \tilde F $ and $ F $, satisfying the boundary conditions
\begin{align}
	\tilde F'(0) = 0 \,, && \tilde F(1) = 1 = F(1) \,.
\end{align}
Note that in the sub-critical case the domain of definition of $ F $ is $ [1, \infty) $, but for super-critical string tensions it is the same as that of $\tilde F $, viz.\ $ [0, 1] $.
We will specify $ \tilde F $ and $ F $ when discussing the solutions below.

This completes the specification of initial data. Indeed, the remaining variables, $ \tilde\eta_{0i} $ and $ \eta_{0i} $ are determined by the regularity condition \eqref{eq:reg_axis_eta} together with the constraints~\eqref{eq:ER_vac_eta_r} and \eqref{eq:00_junc_cond}. Specifically,
\begin{subequations}
\label{eq:eta0Tilde_init}
\begin{align}
	\tilde \eta_{0i} & = \int_0^{R} \!\rd \tilde r \, \tilde r \left[ \left(\partial_{\tilde r} \tilde\alpha_i \right)^2 + \left(\partial_{\tilde t} \tilde\alpha_i \right)^2 \right] \label{eq:eta0Tilde_init_a}\\
	& = \frac{H_i^2}{\gamma_i^2} \int_0^{R} \!\rd \tilde r \, \tilde r \tilde F^2(\tilde r / R) \\
	& = \frac{H_i^2 R^2}{\re^{-2\tilde\eta_{0i}} + H_i^2 R^2}   \int_0^{1} \!\rd x \, x \tilde F^2(x)\, ,
\end{align}	
\end{subequations}
an implicit equation for $ \tilde\eta_{0i} $ which can be solved numerically.
It is also interesting that $ \tilde\eta_{0i} $ is a direct measure of the gravitational energy stored inside the cylinder initially, which is suggested by~\eqref{eq:eta0Tilde_init_a}. In fact, it is (up to a constant factor) nothing but the so called {\it C-energy} introduced by Thorne~\cite{Thorne:1965}.
The exterior $ \eta_{0i} $ is then obtained from~\eqref{eq:00_junc_cond}, evaluated at initial time:
\begin{equation}
	\bar\lambda = \sqrt{ \re^{-2\tilde\eta_{0i}} + H_i^2 R^2 } - \sigma \sqrt{ \re^{-2\eta_{0i}} + H_i^2 R^2 } \, .
\end{equation}
The sign $ \sigma $ again ensures that this has a real solution for $ \eta_{0i} $ in both the sub- and super-critical case.

The complete set of initial data therefore consists of the two parameters $ \bar\lambda $, $ H_i R $ and the two functions $ \tilde F(x), F(x) $, which are all dimensionless.

Finally, we have to impose boundary conditions for $ \tilde \alpha, \alpha $ at the boundary of the domain of integration. In the interior, this is the Neumann boundary condition~\eqref{eq:reg_axis_alpha}. In the exterior, we have to distinguish two cases:

For sub-critical string tensions, the exterior $ r $-domain is actually infinite. But the numerical treatment introduces an artificial boundary at some maximal value $ r_\mathrm{max} $. Since we want to model an empty, infinite exterior, the adequate boundary condition would be an outgoing wave condition. However, it is well known that such a criterion is necessarily non-local in time in the case of cylindrically symmetric waves (see~\cite{Givoli:1991} for a review, and~\cite{Hofmann:2013zea} for a discussion in the context of GR). Therefore, it is computationally quite expensive and so we simply choose the most primitive alternative of making the domain of integration large enough so that the waves which are initially produced near the cylinder cannot reach $ r_\mathrm{max} $ by the end of the numerical simulation.

In the super-critical case, the exterior domain ends at $ r = 0 $. We will enforce the same Neumann boundary condition $ \partial_r \alpha |_{r=0} = 0 $ as at the interior axis in this case, in order to avoid $ \alpha $ becoming singular, which would imply a curvature singularity.

\subsection{Sub-critical tension}

Before turning to the super-critical case in Sec.~\ref{sec:num_supercrit}, let us first present a dynamical solution for a sub-critical tension. This will help us gain confidence in the numerical solver and explicitly demonstrate that the deficit angle geometry reviewed in Sec.~\ref{sec:static} is an attractor solution.
As an example let us choose the parameters
\begin{align}
	\bar\lambda = 0.5 \,, && H_i R = 0.1\,,
\end{align}
and the functions
\begin{align}
	\tilde F(x) = 1, && F (x) = \exp\left[ -\frac{(x - 1)^2}{\delta^2} \right] \,,
\end{align}
with $ \delta = 0.1 $, i.e.\ the initial velocity is localized around the regularized string.

\begin{figure}[t]
	\centering
	\includegraphics[width=0.45\textwidth]{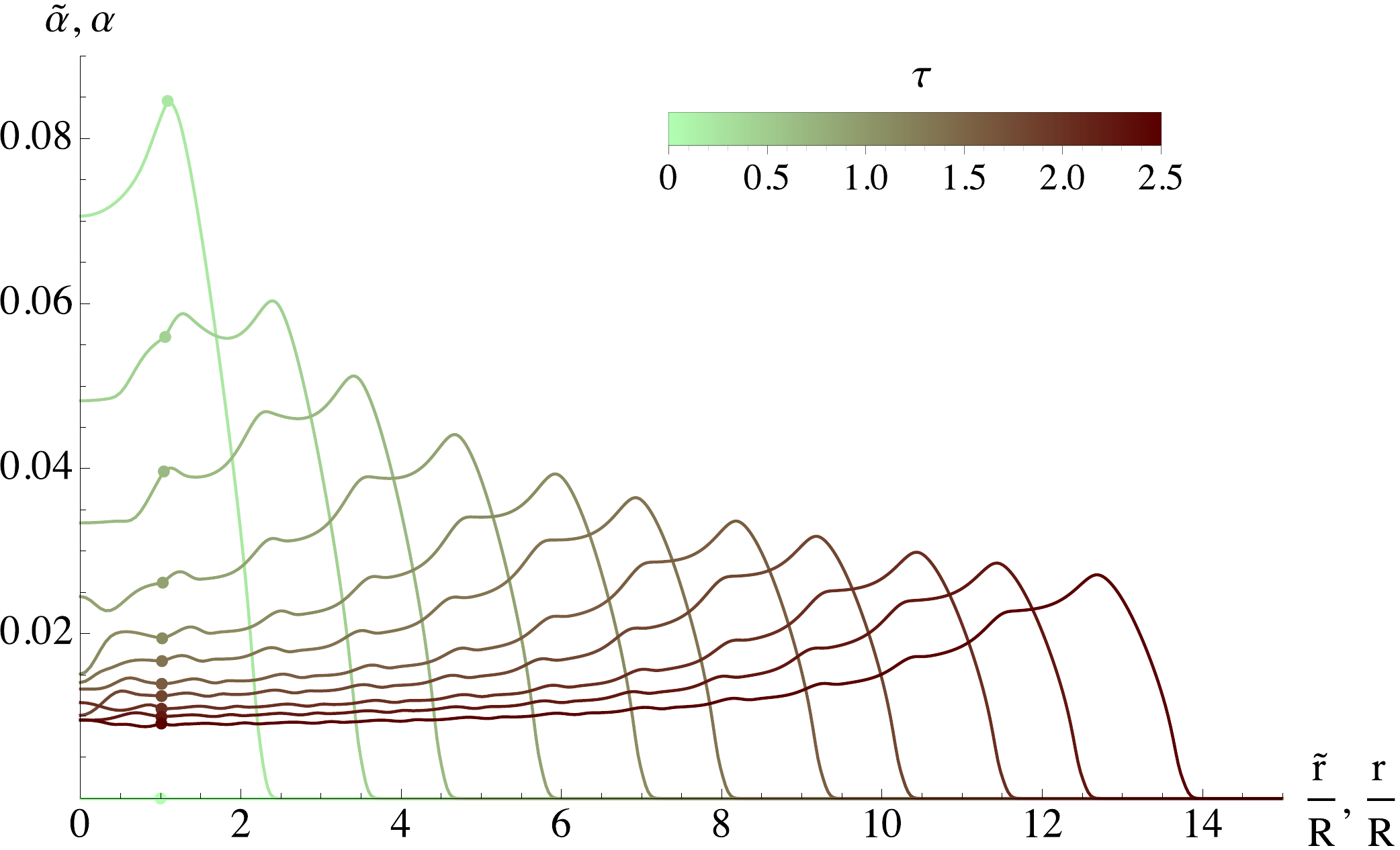}
	\caption{The radial profile of $ \alpha $ at different values of $ \tau $ for a sub-critical tension. The dots indicate the shell's position, left of which the plotted function is $ \tilde\alpha(\tilde r) $. After the initial perturbation is carried away in form of outgoing gravitational waves, the metric settles back to the static deficit angle geometry.}
	\label{fig:alpha_subcrit}
\end{figure}

Fig.~\ref{fig:alpha_subcrit} shows the combined radial profile of the metric function $ \tilde\alpha $ and $ \alpha $ for various moments of time. The initial velocity profile leads to a rapid increase around the position of the cosmic string. Subsequently, it falls back down, thereby emitting cylindrically symmetric gravitational waves. Meanwhile, the coordinate position of the cylinder (indicated as dots in the plots) stays approximately constant. The small oscillations in the $ r $-profile of $ \alpha $ of frequency $ \sim 1/R $ are due to waves in the interior of the cylinder which are reflected at the axis and partially reflected at the cylinder's surface. At late times, $ \alpha $ asymptotically settles back to a constant profile, i.e.\ back to the static deficit angle solution we started with.

\begin{figure}[t]
\subfloat[]{
	\includegraphics[width=0.45\textwidth]{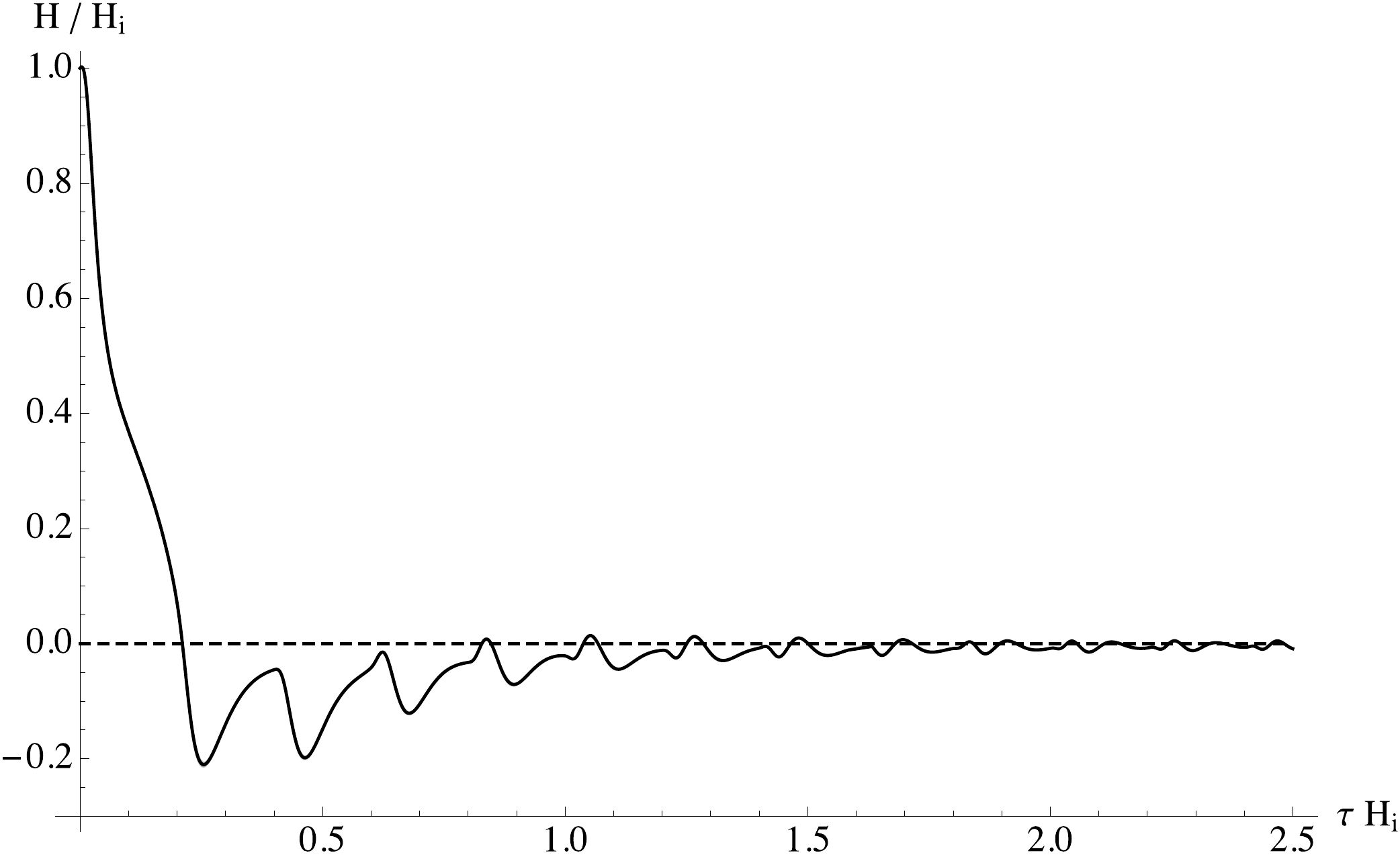}
	\label{fig:hubble_subcrit}
}
\\
\subfloat[]{
	\includegraphics[width=0.45\textwidth]{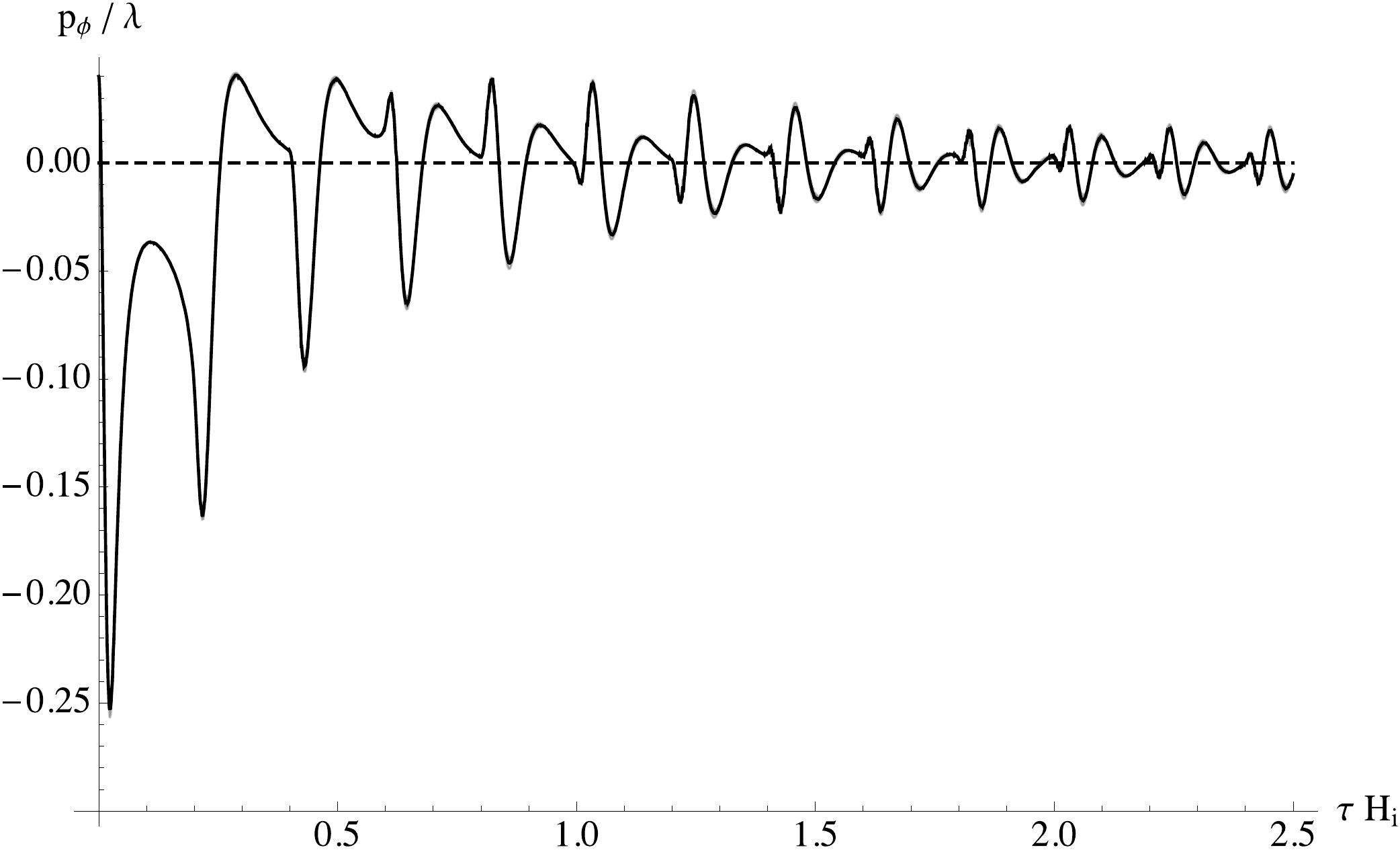}
	\label{fig:wPhi_subcrit}
}
\caption{For sub-critical tensions, both the axial expansion rate $ H $ and azimuthal pressure $ p_\phi $ oscillate and approach zero at late times, in accordance with the analytic predictions for the static solution. The numerical error estimates do not exceed the line thickness.}
\end{figure}

This can also be seen from Fig.~\ref{fig:hubble_subcrit} which shows the expansion rate $ H \equiv \rd \alpha_0 / \rd\tau $ as a function of time. After starting with a positive value, it becomes negative, turns around and asymptotically approaches zero. The oscillatory modulations are again due to the gravitational waves which are moving back and forth in the interior. Finally, Fig.~\ref{fig:wPhi_subcrit} shows the effective equation of state of the stabilizing pressure $ p_\phi $. Again, the oscillatory behavior is imprinted in the evolution; but more importantly, we see that it never becomes smaller than $ -1 $, and therefore $ p_\phi $ is physically reasonable in the sense that it satisfies the Null Energy Condition. Furthermore, at late times it approaches zero, in agreement with the prediction \eqref{eq:static_constants}.

We checked that the qualitatively same behavior is found for other values of $ \bar\lambda $ and $ H_i R $, as long as they satisfy the condition \eqref{eq:sub_crit}: the system always approaches the static conical defect geometry at late times.
Hence, this solution is indeed an attractor in the case of sub-critical string tensions.

\subsection{Super-critical tension} \label{sec:num_supercrit}

Next, let us turn to the actual case of interest --- super-critical string tensions. As an example, we consider the parameters
\begin{align}
	\bar\lambda = 1.5 \,, && H_i R = 0.35 \,,
\end{align}
and a flat initial velocity profile for both\footnote{This choice can be justified a posteriori, because the attractor solutions at asymptotically late times approach roughly constant $ r $-profiles. But we checked that the same attractor solutions are approached for other initial velocity profiles, like e.g.\ a Gaussian as before. Furthermore, they are still approached if $ H_i R $ is made smaller, i.e.\ if the system is perturbed with less energy. 
} $ \tilde \alpha $ and $ \alpha $:
\begin{align} \label{eq:init_profiles_supercrit}
	\tilde F(x) = 1 \,, && F(x) = 1 \,.
\end{align}

\begin{figure}[t]
	\subfloat[]{
		\includegraphics[width=0.45\textwidth]{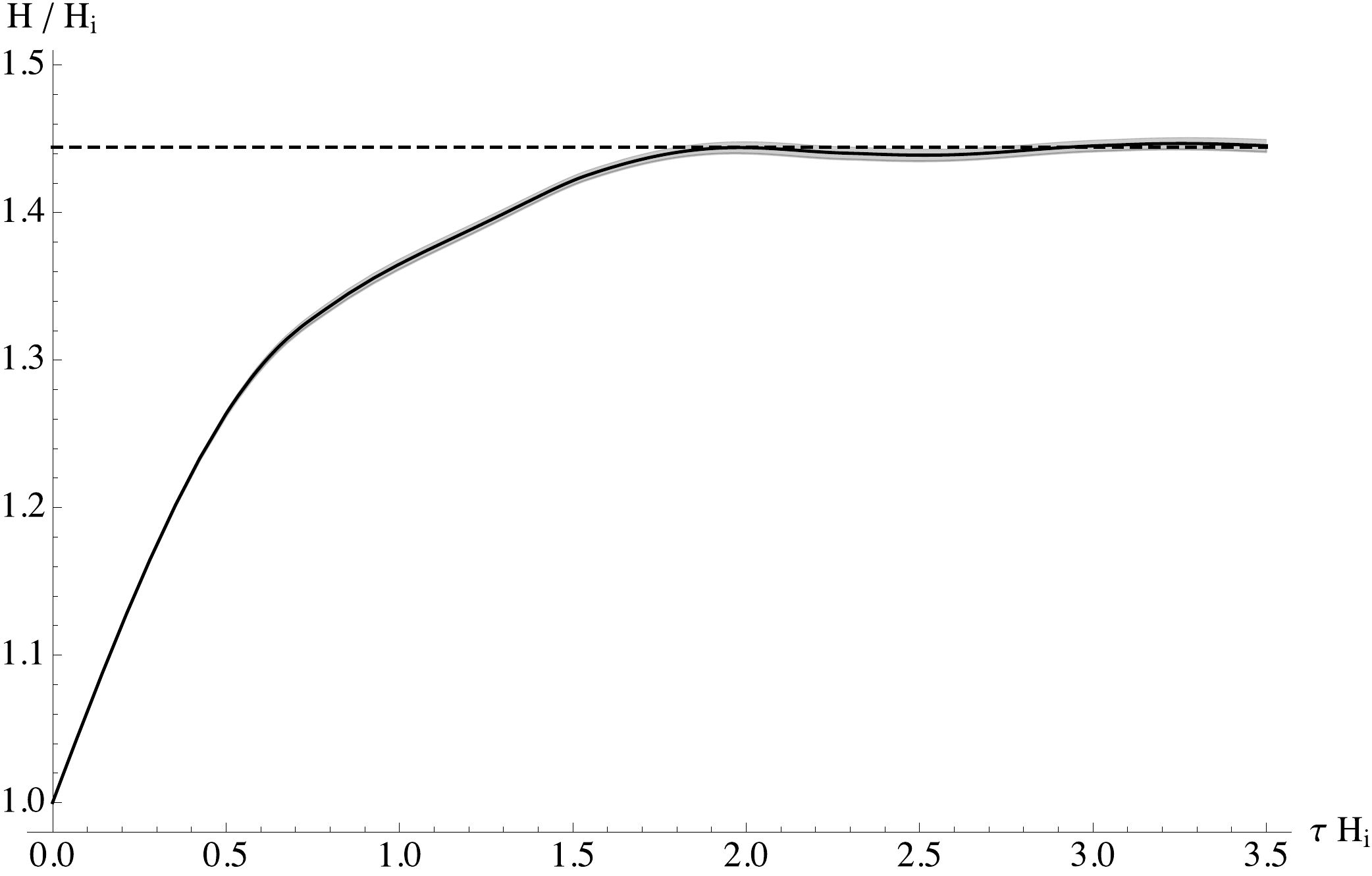}
		\label{fig:hubble_supercrit}
	}\\
	\subfloat[]{
		\includegraphics[width=0.45\textwidth]{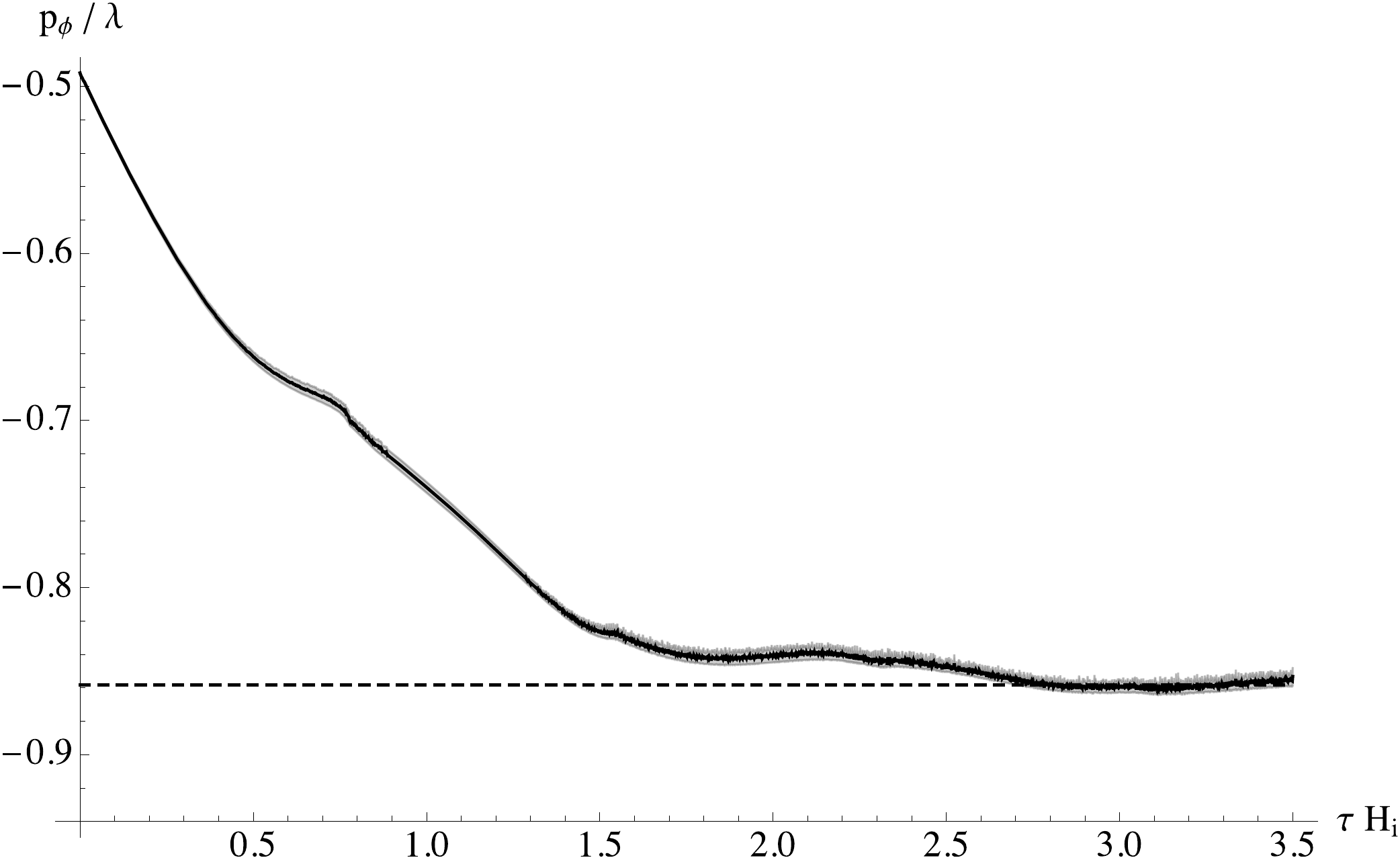}
		\label{fig:wPhi_supercrit}
	}
	\caption{For super-critical tensions, the axial expansion rate $ H $ and azimuthal pressure $ p_\phi $ both tend to constant, non-zero values at late times. This shows that the static solution is not an attractor anymore. The numerical error estimates are indicated by the gray bands. The dashed lines correspond to the analytic predictions derived in Sec.~\ref{sec:analytic}.}
\end{figure}

This time, the system shows a qualitatively completely different behavior. The expansion rate $ H $, depicted in Fig.~\ref{fig:hubble_supercrit}, instead of going to zero, approaches a constant non-zero value. This is one of the main results of the present work: \textit{The static defect angle geometry is no stable solution in the case of super-critical string tensions. Instead, the attractor solutions are those in which the string expands in axial direction at a constant rate.}

The equation of state of the azimuthal pressure is shown in Fig.~\ref{fig:wPhi_supercrit}. It also approaches a constant value at late times; but more importantly, it is again always larger than $ -1 $ and hence consistent with a radial stabilization by means of physically reasonable matter. However, the asymptotic value of $ p_\phi / \lambda $ depends on the tension, as will be discussed in Sec.~\ref{sec:analytic}, which will ultimately lead to a break down of stabilizability.

\begin{figure}[t]
	\centering
	\subfloat[interior]{
		\includegraphics[height=0.3\textwidth]{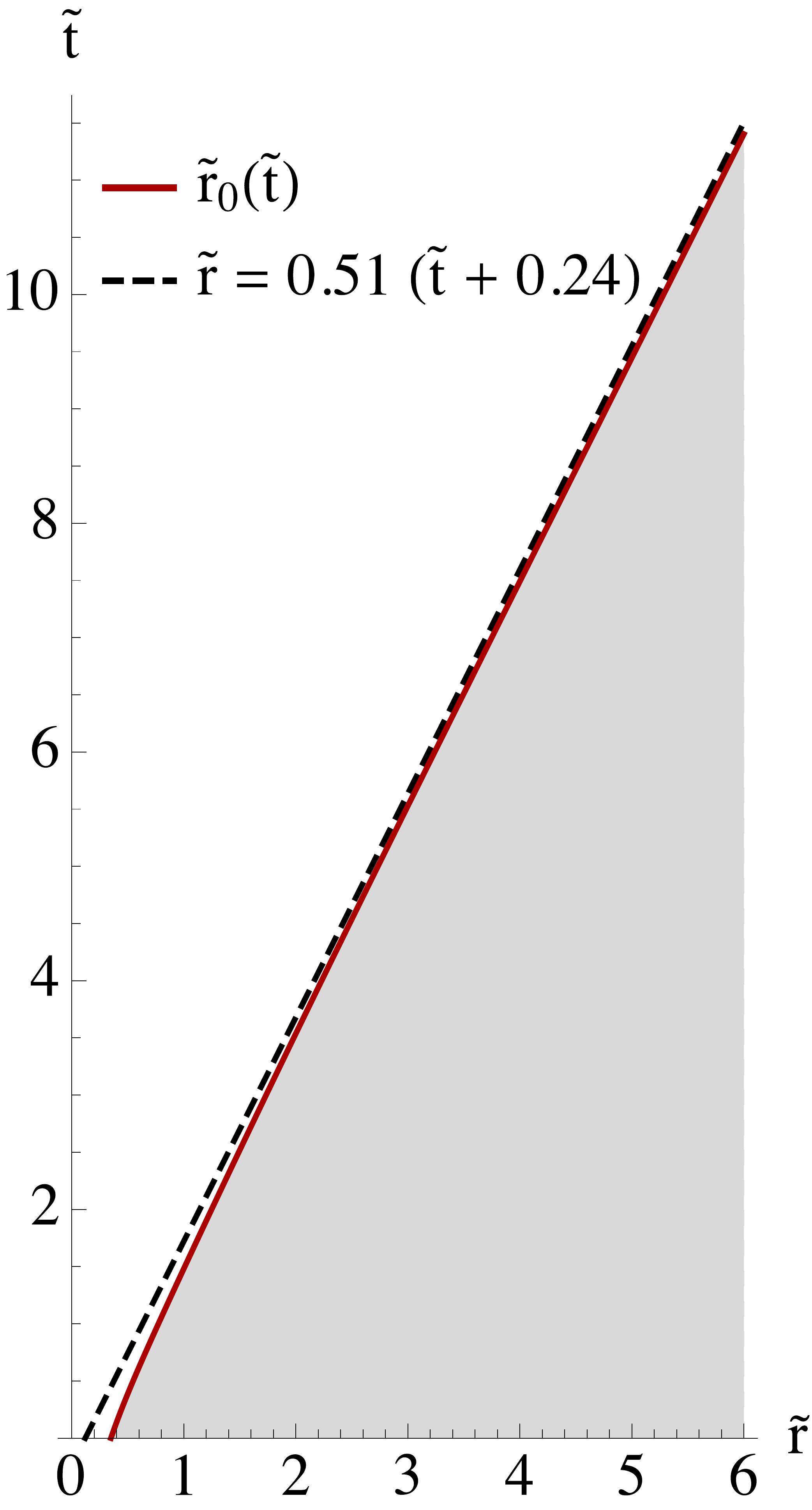}
		\label{fig:r0In}
	}
	\hfill
	\subfloat[exterior]{
		\includegraphics[height=0.3\textwidth]{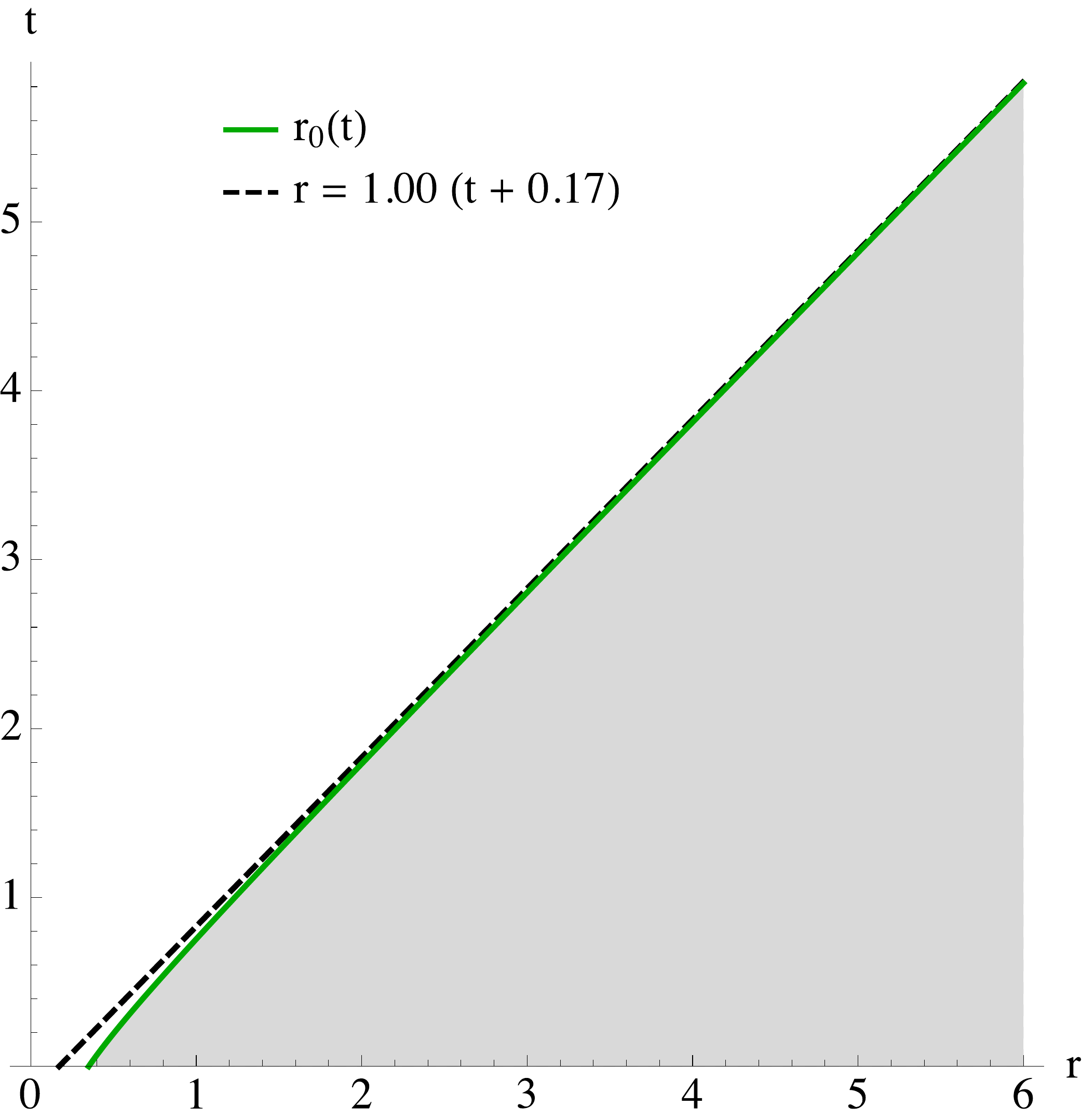}
		\label{fig:r0Out}
	}
	\caption{For super-critical string tensions, the shell's position approaches a constant velocity in the interior and in the exterior coordinate patch. In the interior, this asymptotic velocity is $ < 1 $ and its actual value depends on $ \bar\lambda $. In the exterior, it is generically $ = 1 $, implying a horizon. The coordinates in either patch only range from the axes to the shell, so the gray regions are not part of the spacetime.}
	\label{fig:supercrit_r0}
\end{figure}

The radial coordinate position of the cylinder is now no longer approximately constant, but approaches a constant velocity, as can be seen in Fig.~\ref{fig:supercrit_r0}. Quite remarkably, it turns out that in the exterior coordinate patch, this asymptotic velocity is $ 1 $; this is just the speed of light, since in the coordinates~\eqref{eq:met_ER} radial light rays correspond to $ \rd r = \pm \rd t $. This means that no signal from beyond the dashed line in Fig.~\ref{fig:r0Out} can ever reach the string, drawn as a solid (green) line, or in other words:
\textit{A horizon is formed  outside the super-critical cosmic string.}
This is the second main result of our analysis.

On the other hand, the asymptotic velocity in the interior coordinate patch is less than $ 1 $, so no horizon is formed inside the regularized string.
Note that, even though the exterior speed asymptotically approaches unity, it always stays below $ 1 $. Otherwise, there would also be contradictions because: (i) the shell represents a massive object, which can not travel exactly at the speed of light; (ii) moving at the speed of light is a coordinate invariant statement, so if it did hold in the exterior, it would also have to hold in the interior coordinate patch.

In the example we showed in Fig.~\ref{fig:supercrit_r0}, the conical singularity at $ r=0 $, from the string's point of view, is hidden behind the horizon.
However, this is no generic feature of the solutions, because the actual position of the horizon depends on the initial condition $ H_iR $: smaller values move the dashed line in Fig.~\ref{fig:r0Out} to the left. By choosing $ H_i R $ small enough, the horizon can be pushed so far that it crosses the axis, implying that the conical singularity is no longer hidden. But as will be shown in Sec.~\ref{sec:remove_conical_sing}, for time dependent geometries the conical singularity can be avoided altogether.

\subsection{Radial geometry} \label{sec:radial_geom}

\begin{figure*}[htb]
	\subfloat[Sub-critical ($ \bar\lambda = 0.5 $): After the Einstein-Rosen waves have been emitted, the geometry asymptotically settles to the static defect angle solution.]{
		\includegraphics[width=0.19\textwidth]{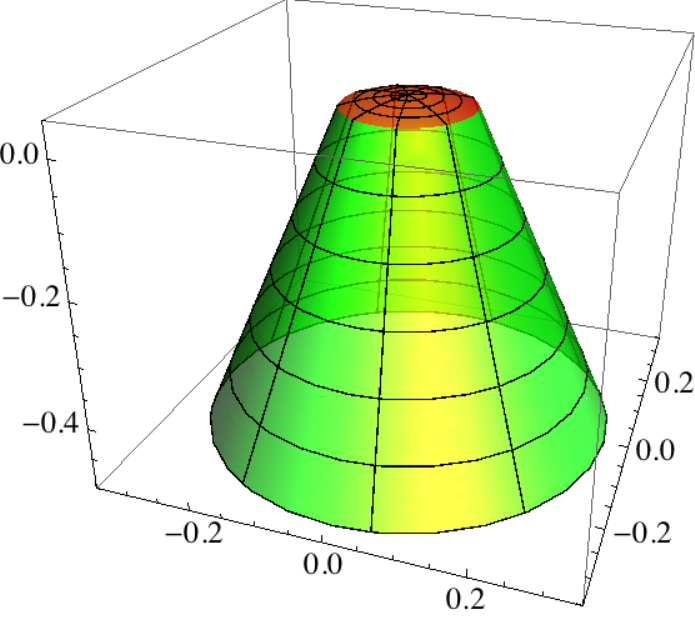}\hfill
		\includegraphics[width=0.19\textwidth]{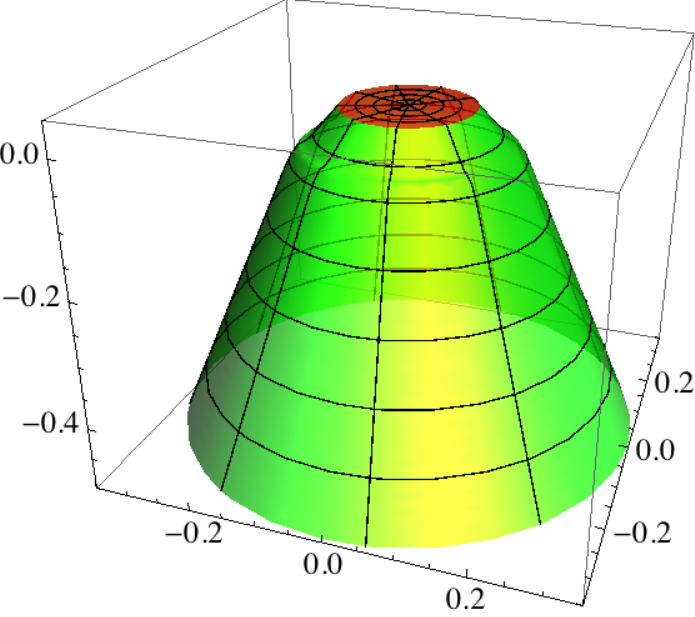}\hfill
		\includegraphics[width=0.19\textwidth]{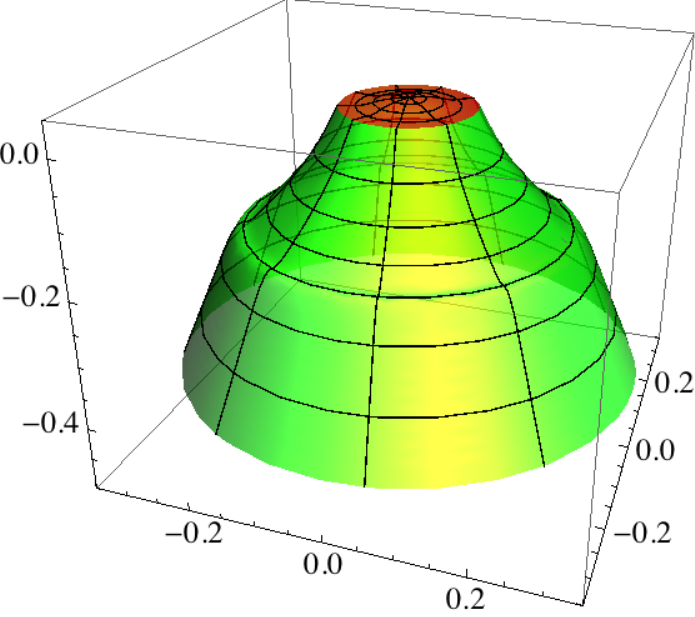}\hfill
		\includegraphics[width=0.19\textwidth]{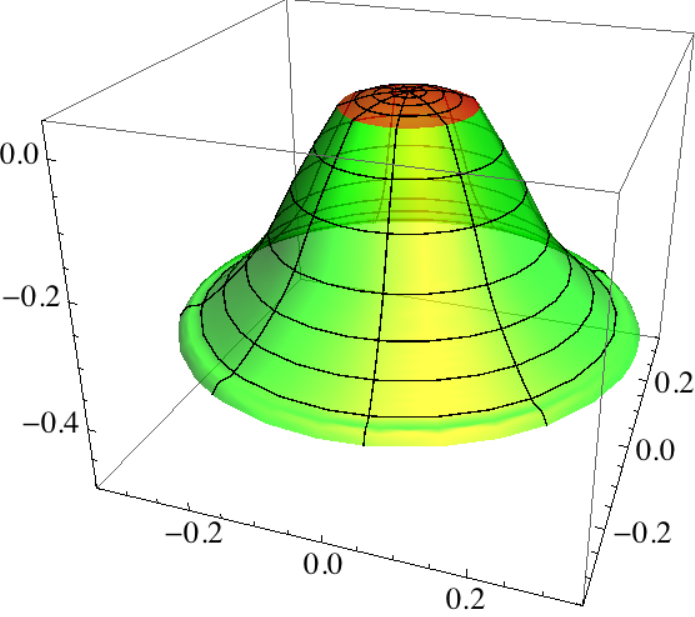}\hfill
		\includegraphics[width=0.19\textwidth]{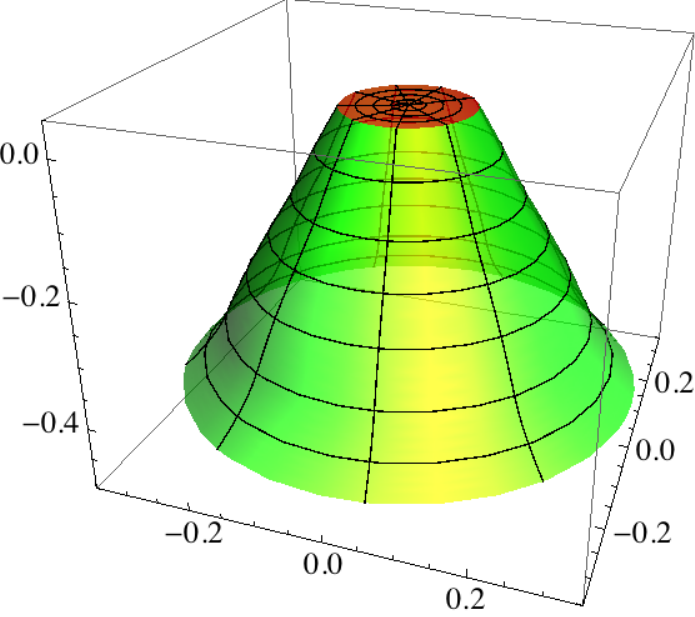}
		\label{fig:embed_subcrit}
	}%
\\
	\subfloat[Super-critical ($ \bar\lambda = 1.5 $): The interior initially oscillates and settles to a constant profile at late times. The exterior keeps growing, and the light ray stays at a finite distance away from the shell, due to the horizon.]{
		\includegraphics[width=0.19\textwidth]{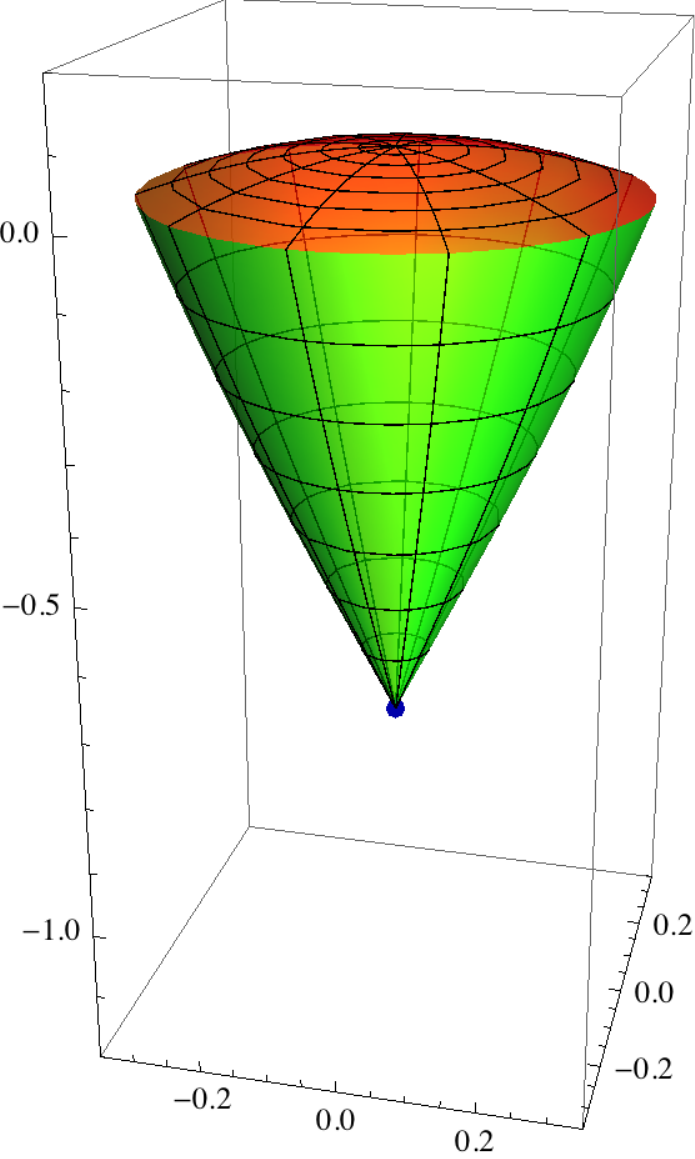}\hfill
		\includegraphics[width=0.19\textwidth]{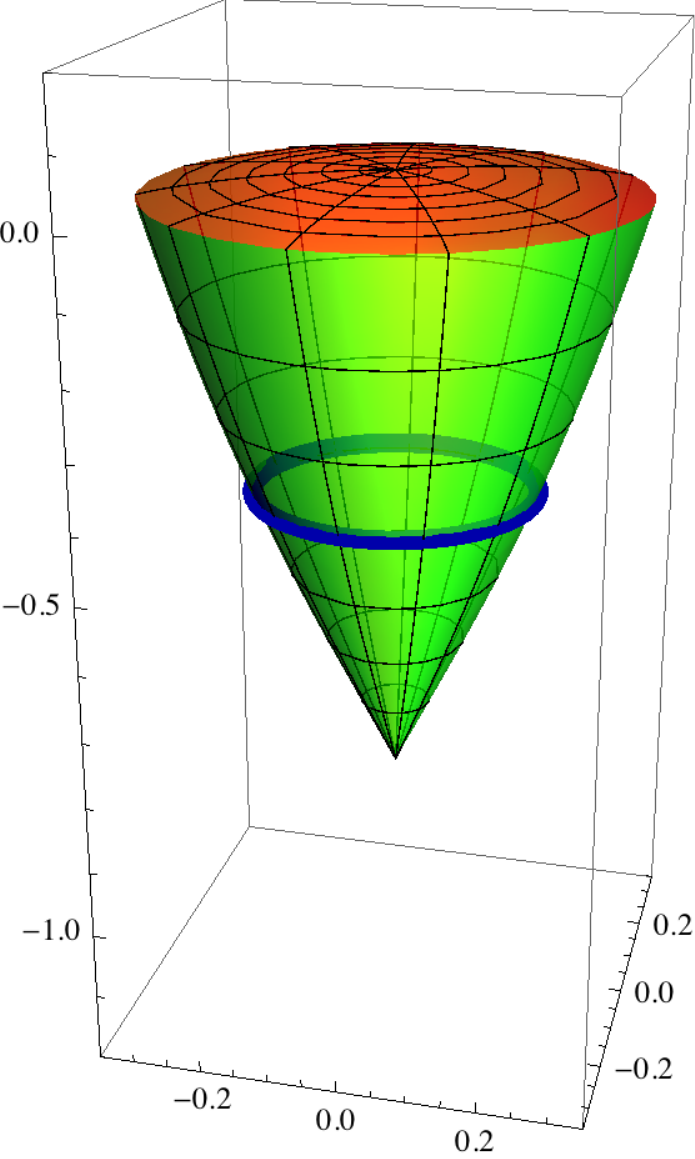}\hfill
		\includegraphics[width=0.19\textwidth]{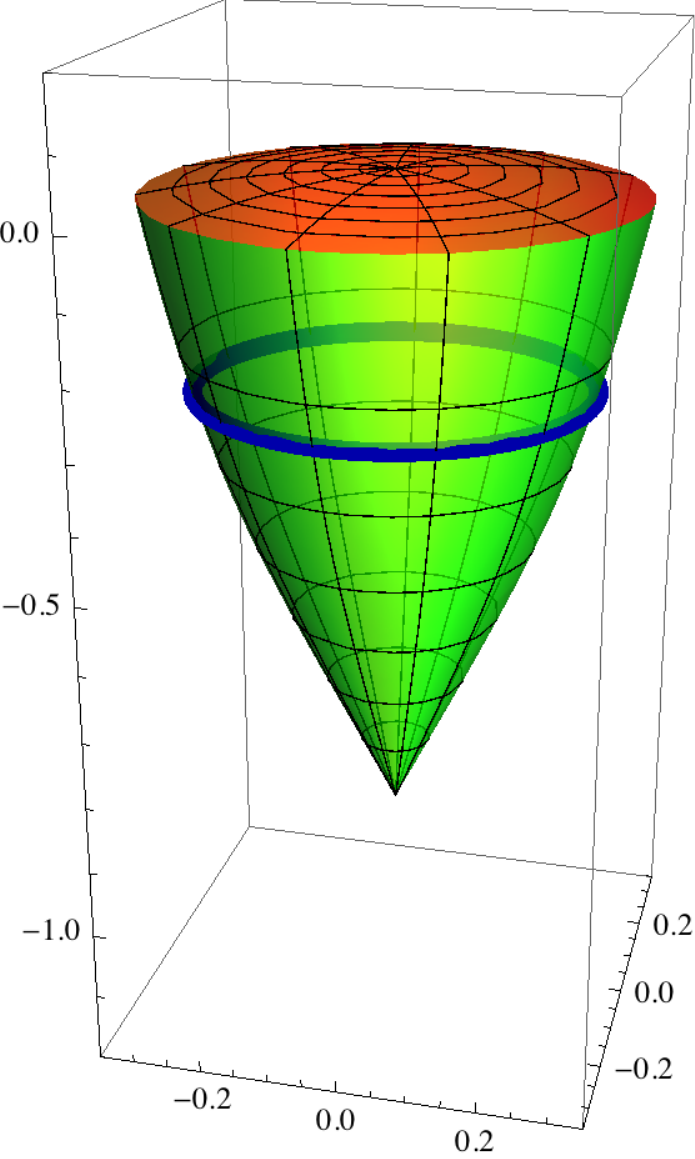}\hfill
		\includegraphics[width=0.19\textwidth]{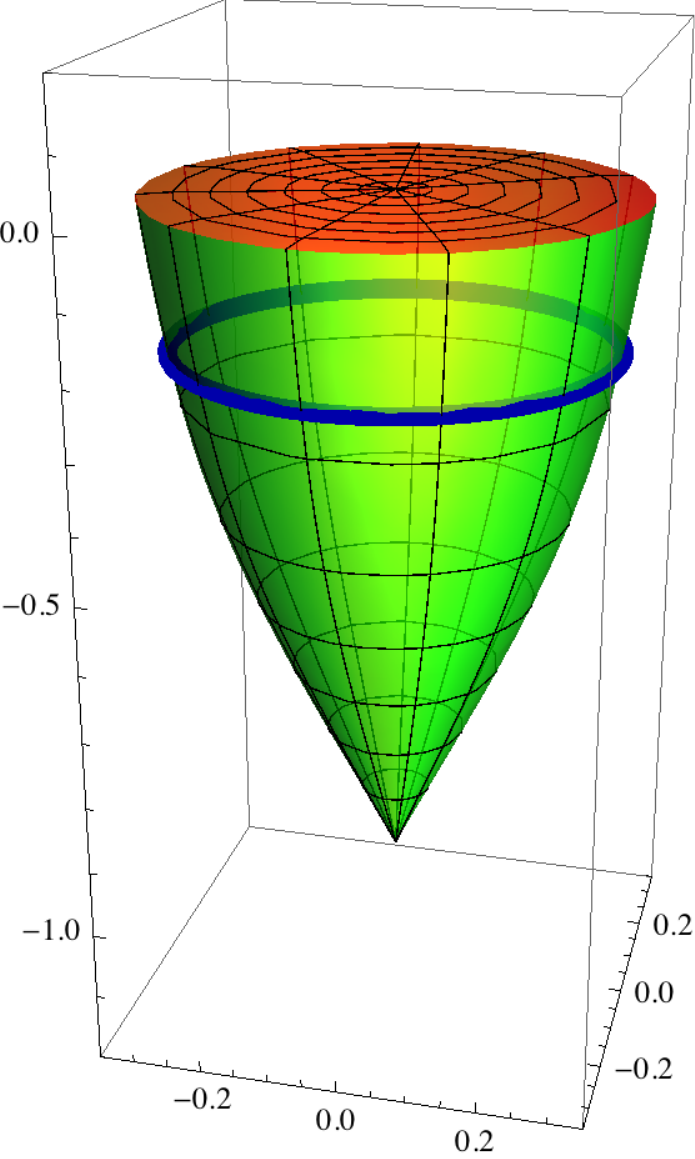}\hfill
		\includegraphics[width=0.19\textwidth]{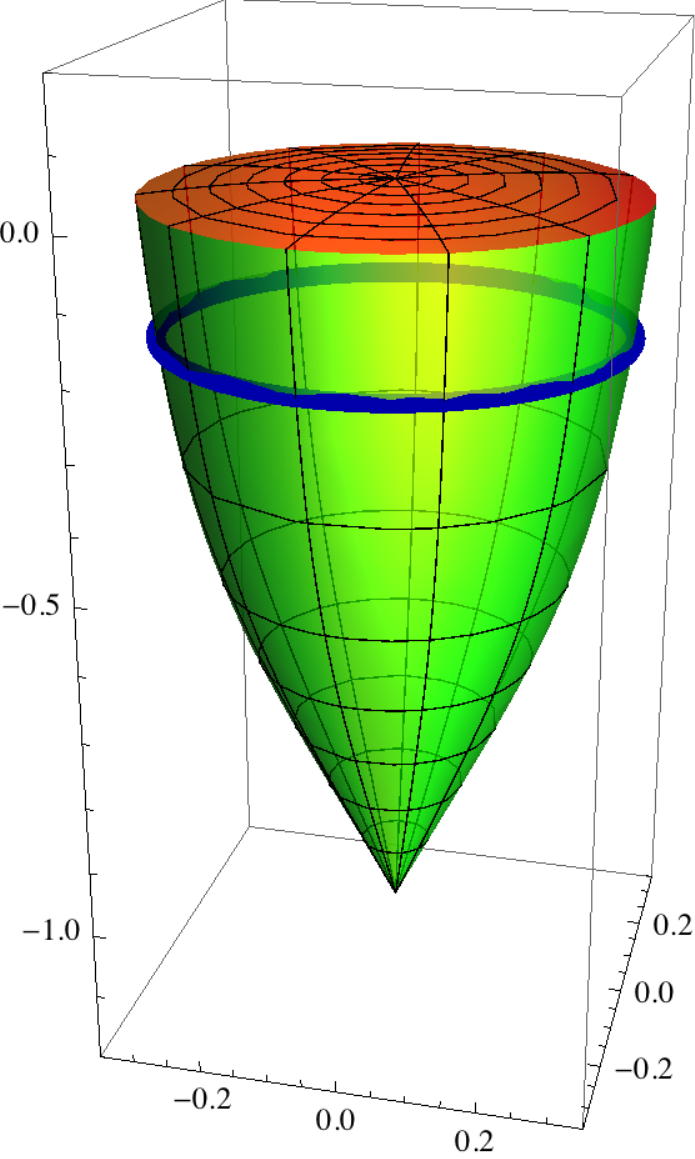}
		\label{fig:embed_supercrit}
	}%
\\
	\subfloat[Super-critical ($ \bar\lambda = 1.5 $), without a conical singularity. The qualitative behavior is the same as before.]{
		\includegraphics[width=0.19\textwidth]{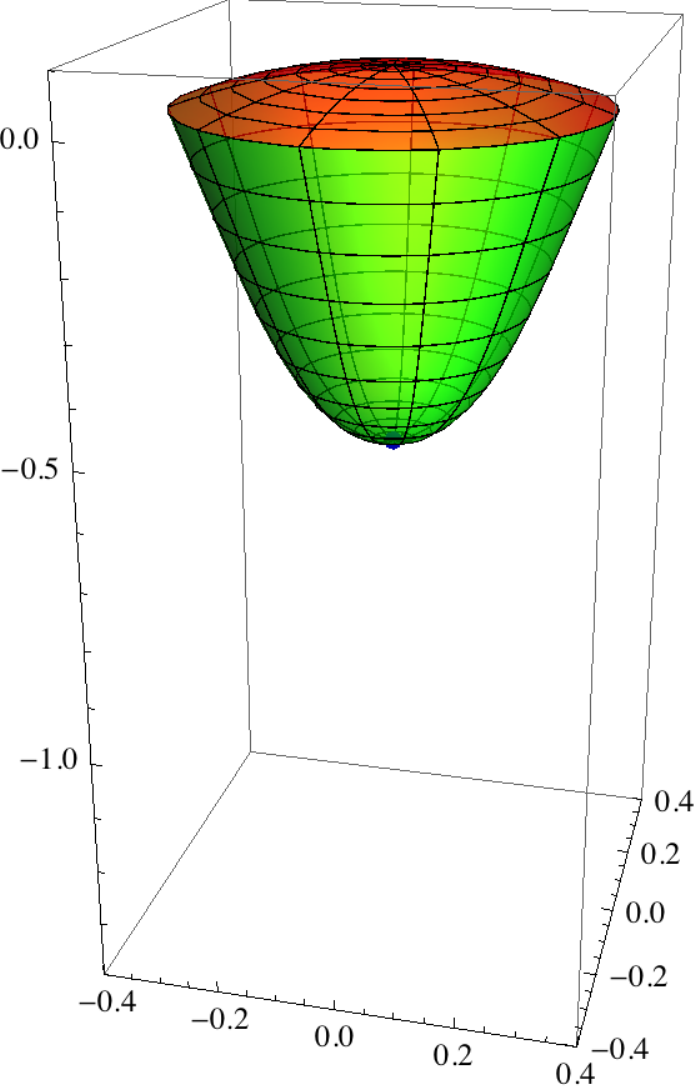}\hfill
		\includegraphics[width=0.19\textwidth]{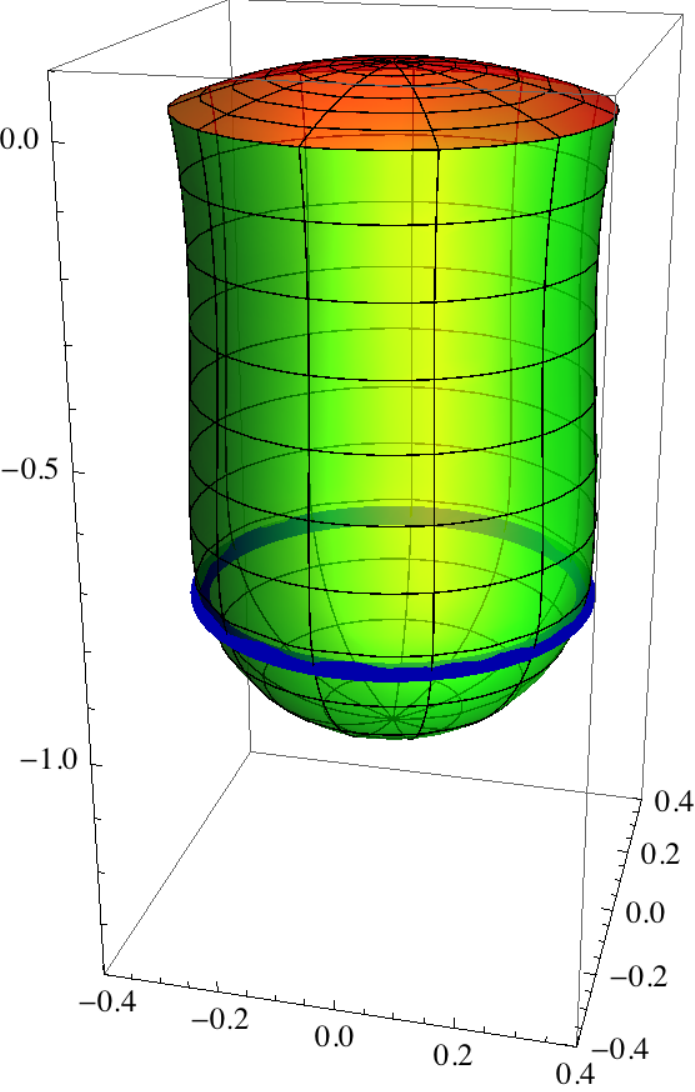}\hfill
		\includegraphics[width=0.19\textwidth]{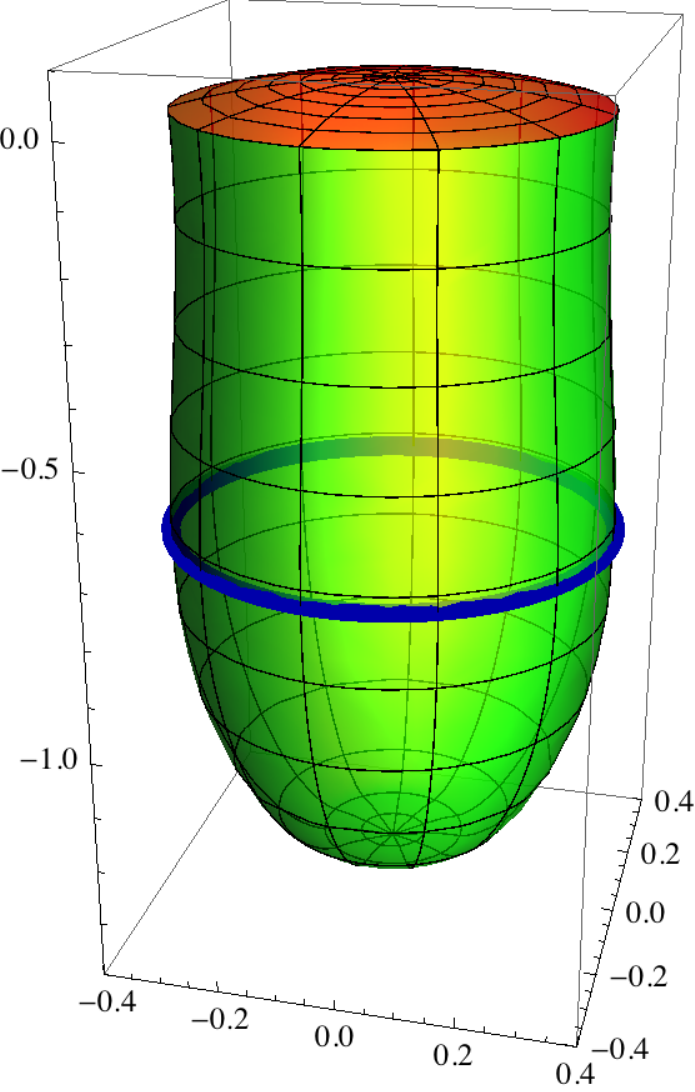}\hfill
		\includegraphics[width=0.19\textwidth]{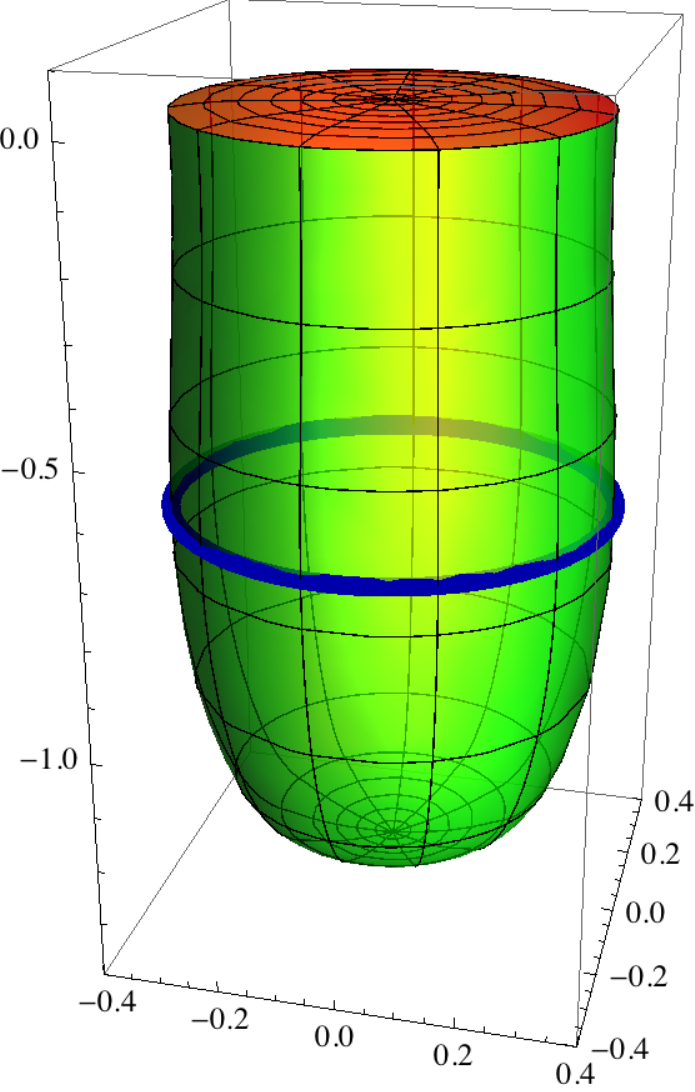}\hfill
		\includegraphics[width=0.19\textwidth]{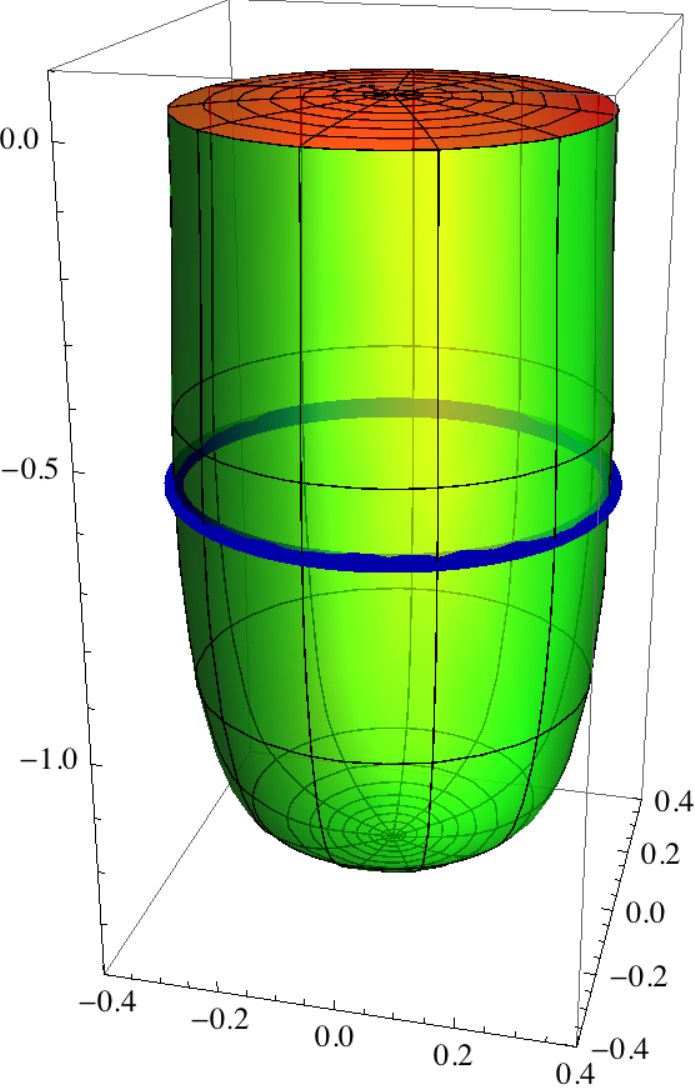}
		\label{fig:embed_supercrit_reg}
	}%
	\caption{(Colors online) Embedding diagrams of the radial geometry at equidistant time steps $ \Delta\tau $. The interior of the regularized string is drawn red (dark gray), the exterior green (light gray). In the super-critical cases, a light ray emitted from the exterior axis towards the shell is drawn as a solid dark line, visualizing the formation of a horizon.} \label{fig:embed}
\end{figure*}

\begin{figure*}[htb]
	\subfloat[]{
	\includegraphics[width=0.45\textwidth]{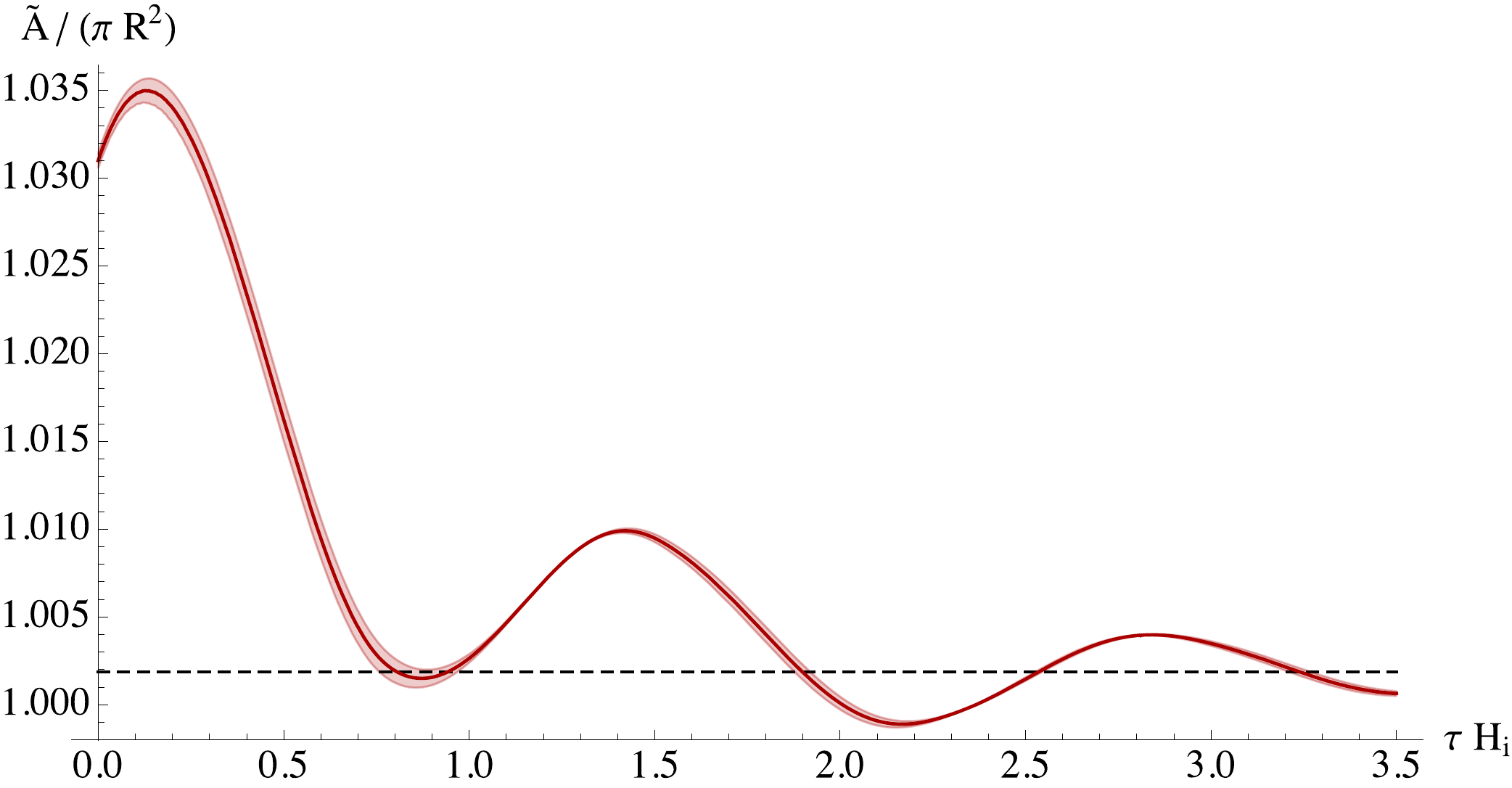}
	\label{fig:volIn_supercrit}
	}
	\hfill
	\subfloat[]{
	\includegraphics[width=0.45\textwidth]{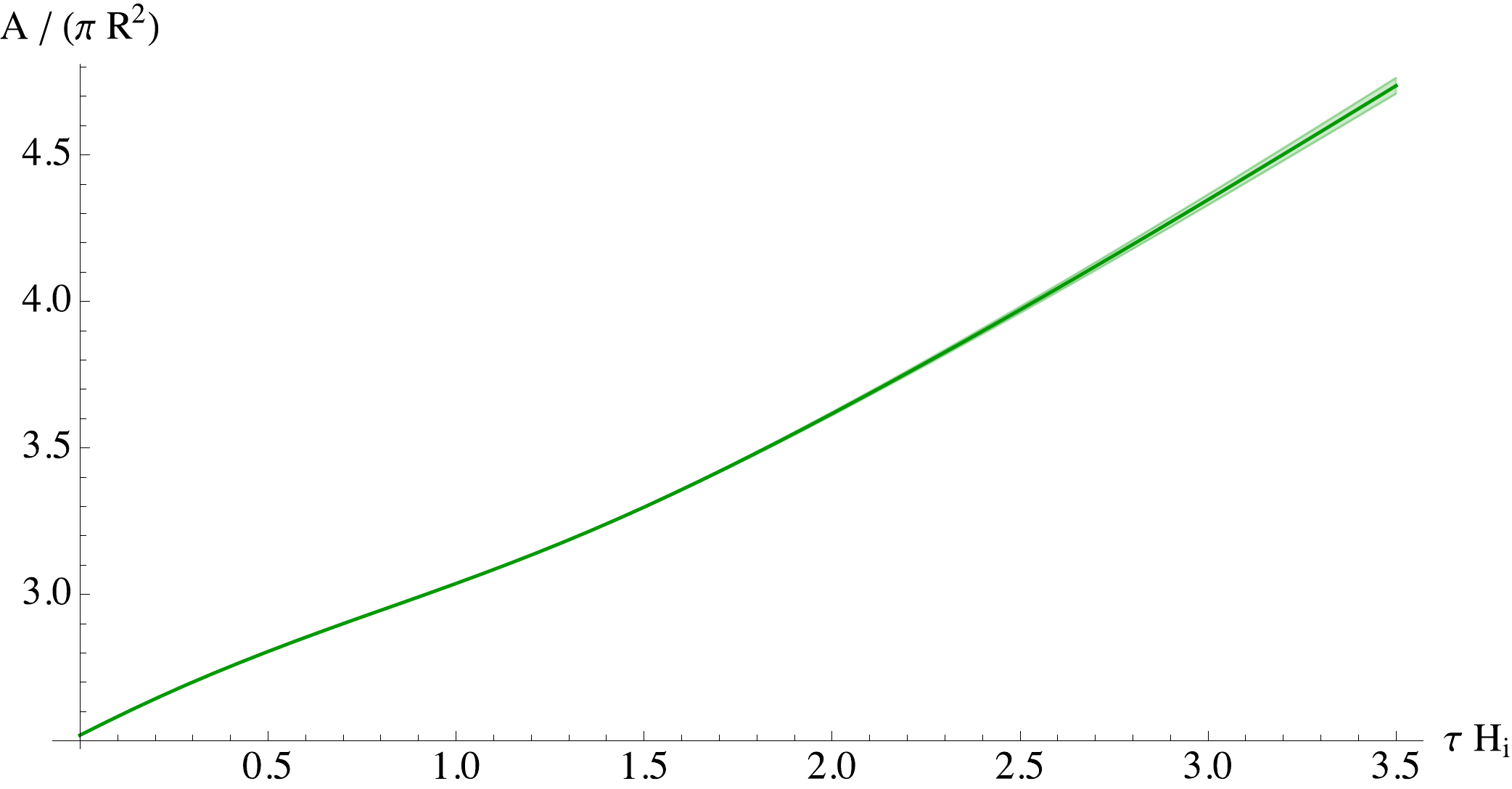}
	\label{fig:volOut_supercrit}
	}\\
	\subfloat[]{
	\includegraphics[width=0.45\textwidth]{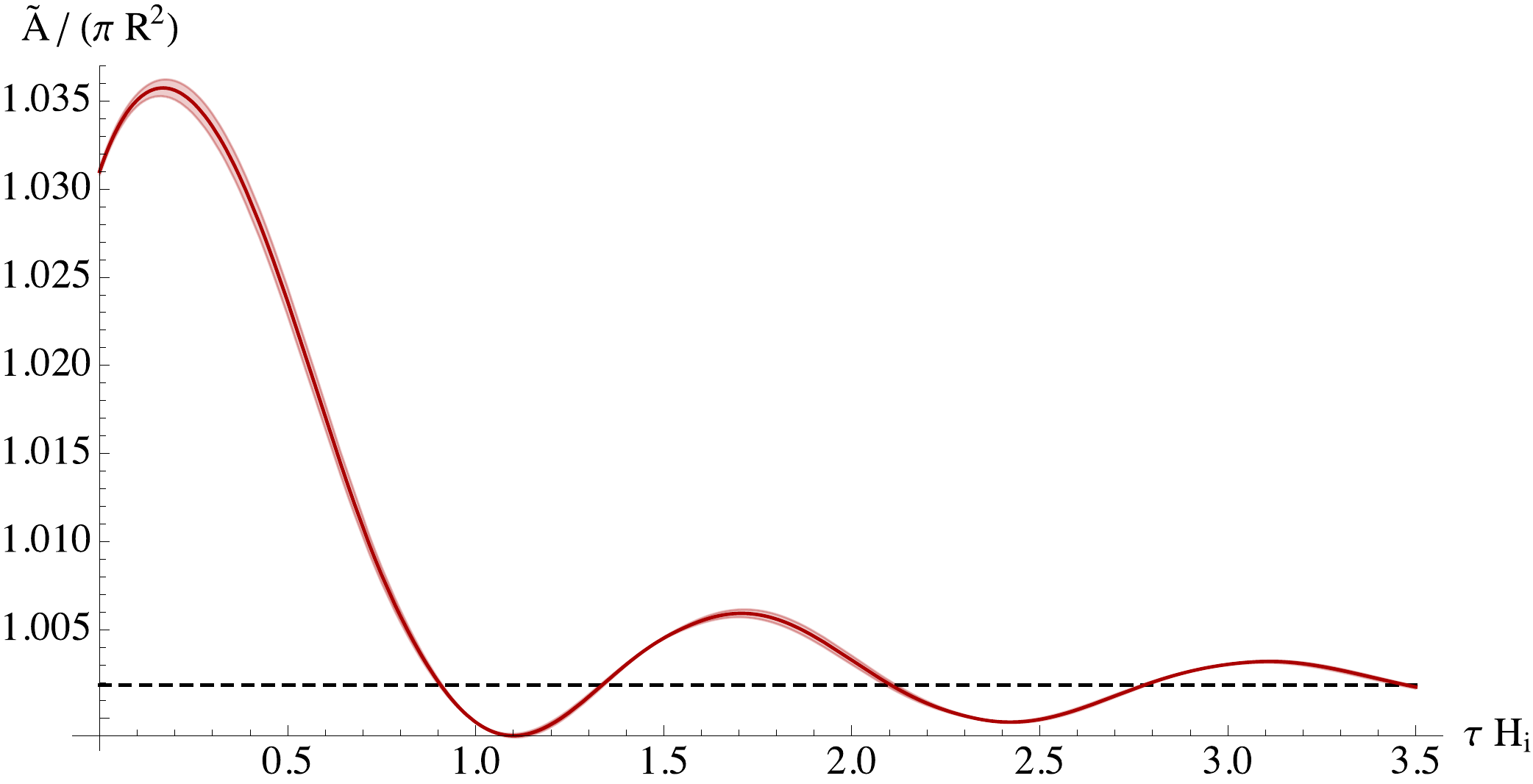}
	\label{fig:volIn_supercrit_reg}
	}
	\hfill
	\subfloat[]{
	\includegraphics[width=0.45\textwidth]{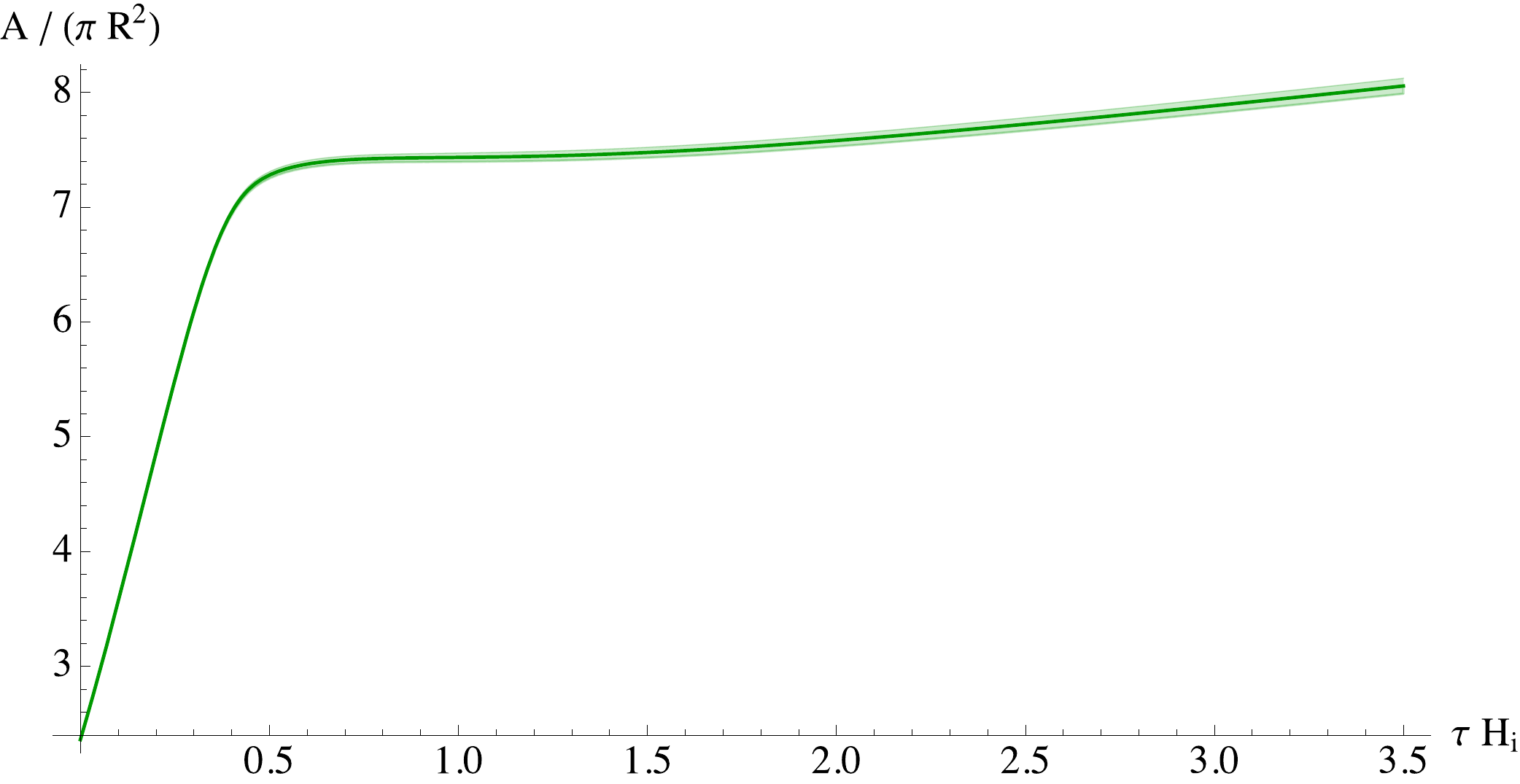}
	\label{fig:volOut_supercrit_reg}
	}
	\caption{Interior (left column) and exterior (right column) area of the super-critical cosmic string geometry with (upper row) and without (lower row) conical singularity at the exterior axis. The interior area approaches a constant value, confirming a successful stabilization, whereas the exterior area keeps growing.} \label{fig:volume}
\end{figure*}

As already mentioned, the intrinsic geometry on the cylindrical shell is fully characterized by the axial expansion rate $ H $, because after fixing $ R $ this expansion is the only non-trivial feature of the induced metric~\eqref{eq:induced_met}. However, the full metric~\eqref{eq:met_ER} contains much more information, namely how space inside and outside the shell is curved in the radial direction, and how it evolves in time. The interior is of particular interest because we want to model a cosmic string with no strong dynamics inside, and so we would like the interior cross-sectional area to be approximately constant in order to have a successful regularization. Fixing $ R $ only keeps the shell's circumference constant, but does not a priori say anything about the area. The constant velocity of the shell in the interior could even lead one to suspect that the area might actually be growing. However, the constancy of the velocity is of course a coordinate dependent statement, so one has to look at the invariant\footnote{It is still slicing dependent, but this does not affect the question whether the area is asymptotically constant or growing.} area. Furthermore, it would be nice to get some intuition about what the exterior geometry actually ``looks like''.

The radial warping of the full metric can be visualized by drawing embedding diagrams of the two-dimensional temporal and axial slices (i.e.\ $ t, z =$ constant) in a fictitious three-dimensional space. These diagrams are shown in Fig.~\ref{fig:embed}, where the interior is drawn red (dark gray) and the exterior green (light gray). The series of diagrams in each row corresponds to snapshots taken at equidistant times $ \tau $, starting at $ \tau = 0 $ on the left and increasing to the right.

Fig.~\ref{fig:embed_subcrit} corresponds to the sub-critical case $ \bar\lambda = 0.5 $. The exterior space is cut off at some finite radius in the pictures, but actually extends to infinity. One can clearly see how the shell creates a deficit angle, making the exterior space conical. As time evolves, the disturbance induced by the initial conditions moves outwards in the form of a cylindrical gravitational wave, and the geometry asymptotically settles to the static defect angle solution\footnote{The initial defect angle is slightly larger than the final one, because initially the non-zero value of $ H $ gives an additional contribution.}.

The super-critical case $ \bar\lambda = 1.5 $ is shown in Fig.~\ref{fig:embed_supercrit}. Here, the exterior space closes up and ends in a conical singularity, making the space compact. The causal structure is visualized by adding a radial light ray that is initially emitted from the exterior axis (solid blue lines). Asymptotically, it stays at a finite distance away from the shell, in accordance with the formation of a horizon.
Furthermore, one can already see from these diagrams that the interior area indeed stays approximately constant, in accordance with a successful stabilization, whereas the exterior area increases. To make these statements more quantitative, Fig.~\ref{fig:volIn_supercrit} shows the interior 2D area
\begin{equation}\label{eq:area}
	\tilde A(\tau) := 2\pi \int_0^{\tilde r_0(\tau)} \tilde r \re^{\tilde\eta - 2\tilde\alpha} \,\rd \tilde r \,.
\end{equation}
The gravitational waves moving back and forth inside the shell lead to very small oscillations in this area, but at late times it indeed approaches a constant value. In Sec.~\ref{sec:analytic} we will derive an analytic prediction for its value, which is shown as a dashed line. Hence, our numerical solution can indeed be viewed as a successful regularization of a stabilized cosmic string. On the other hand, the exterior area $ A(\tau) $, plotted in Fig.~\ref{fig:volOut_supercrit}, gets larger at an asymptotically constant rate.

Finally, Fig.~\ref{fig:embed_supercrit_reg} shows the embedding geometry for the same super-critical tension $ \bar\lambda = 1.5 $, but with initial conditions which remove the conical singularity, as discussed in Sec.~\ref{sec:remove_conical_sing}. These non-trivial initial conditions lead to a much more rapid increase in the exterior area for the first time steps. But apart from that, the qualitative behavior is the same as in the case with conical singularity. In particular, the interior area again approaches the constant value and the exterior size keeps growing, see Fig.~\ref{fig:volIn_supercrit_reg} and~\ref{fig:volOut_supercrit_reg}. However, the asymptotic growth rate is smaller than in the conical case, showing that the speed at which the exterior space gets larger is not only set by the string tension, but also influenced by the amount of C-energy that is needed to smooth out the conical singularity.
\section{Analytic results} \label{sec:analytic}

From the numerical investigations we learned that for super-critical string tensions, the system asymptotically approaches a constant axial expansion rate $ H $ at late times.
In this section, we will derive the analytic relation between the tension $ \bar\lambda $ and $ H R $.
To this end, we first of all make use of the fact that the numerical results reveal another quite generic behavior: For different choices of initial conditions the shell generically approaches a constant (coordinate) velocity 
\begin{align}\label{eq:limit_attract}
\dot {\tilde r}_0(\tilde t) \rightarrow \tilde v < 1 \qquad \text{and} \qquad \dot { r}_0(t) \rightarrow 1\;.
\end{align}
Plugging this into \eqref{eq:r0Dot_rel}, we conclude that $\tilde \gamma$ and $\gamma$ also approach constants\footnote{Using \eqref{eq:gamma_rel} to eliminate $\gamma$ in \eqref{eq:00_junc_cond} shows that the solution asymptotically approaches the critical bound~\eqref{eq:crit} from above.}:
\begin{align} \label{eq:gamma_rel}
	\tilde\gamma \to \frac{H R}{\tilde v} \,, && \gamma \to H R \,. 
\end{align}
Substituting this into the junction condition \eqref{eq:00_junc_cond}, we obtain
\begin{empheq}[box=\widefbox]{align}
\label{eq:system1}
	\bar \lambda = H R \left(1+\frac{1}{\tilde v} \right)\;.
\end{empheq}
This is not yet the relation we are looking for, because it still contains the additional unknown parameter $ \tilde v $. But the numerics show that the value of $\tilde v$ only depends on $ \bar\lambda $ (and not on the initial conditions). Hence, there should be a second relation between the parameters lifting the degeneracy of~\eqref{eq:system1}. However, in order to derive this relation analytically, one needs to know the complete interior geometry of the attractor solution. Fortunately, it turns out that this solution can indeed be found.

We look for a solution $\tilde \alpha$ which leads to a shell coordinate which is changing with a constant rate $\tilde v$, i.e.\ $ \tilde r_0 = \tilde v \tilde t $. (For simplicity, we assume that initially $ \tilde r_0 = 0 $. The general case with an initial offset, which has to be used when comparing to the numerical solutions, is simply obtained by letting $ \tilde t \to \tilde t + \text{const.} $)
Together with~\eqref{eq:defR}, this condition fixes the time dependence of $\tilde \alpha_0$ :
\begin{align}\label{eq:alpha0}
	\tilde \alpha_0 = \ln{\left(\frac{\tilde v \tilde t}{R}\right)}\;,
\end{align}
In order to extend this function into the interior space of the shell, we look for scaling solutions of \eqref{eq:ER_vac_alpha} which depend on $\tilde r$ only through the ratio $ x := \tilde r / \tilde t $. The only solution of this type, which is also compatible with~\eqref{eq:alpha0}, is
\begin{align}\label{eq:alpha_scal}
	\tilde \alpha (\tilde t, \tilde r)= \ln \left[ \frac{\tilde t}{\Omega} \left( 1+\sqrt{1-x^2} \right) \right] \;,
\end{align}
with 
\begin{align}\label{eq:alpha_scaling}
	\Omega := \frac{R}{\tilde v} \left(1+\sqrt{1-\tilde v^2} \right)\;.
\end{align}
Integrating \eqref{eq:ER_vac_eta_r} and \eqref{eq:ER_vac_eta_t} then yields
\begin{align}\label{eq:eta_scal}
	\tilde \eta(\tilde t, \tilde r) = 2 \ln\left(\frac{ 1+\sqrt{1-x^2} }{2\sqrt{1-x^2}}\right),
\end{align}
where the elementary flatness condition $\tilde \eta|_{\tilde r = 0} = 0$ was implemented. 
The complete scaling solution for the interior region now reads
\begin{multline}\label{eq:scaling_sol}
	\rd \tilde s^2 = \left( 1 + \sqrt{1 - x^2} \right)^2 \left[ \frac{\Omega^2}{4} \left(\frac{-\rd \tilde t^2 + \rd \tilde r^2}{\tilde t^2 - \tilde r^2} \right) + \frac{\tilde t^2}{\Omega^2} \rd z^2 \right] \\
	+ \left( \frac{\Omega x}{ 1 + \sqrt{1 - x^2}} \right)^2 \rd \phi^2 \,,
\end{multline}
which is an exact vacuum solution.

We can now evaluate~\eqref{eq:eta_scal} at the shell, use~\eqref{eq:gamma} and~\eqref{eq:gamma_rel} to finally obtain
\begin{empheq}[box=\widefbox]{align}
\label{eq:system2}
	HR = \frac{4 \tilde v\sqrt{1-\tilde v^2}}{\left(1+\sqrt{1-\tilde v^2}\right)^2}\;.
\end{empheq}
This is the second relation we were looking for\footnote{Note that, even though it relates local shell parameters, it implicitly depends on the entire interior geometry through the use of~\eqref{eq:eta_scal}; for instance, it knows about the regularity at the axis.}.
{\it The two equations~\eqref{eq:system1} and \eqref{eq:system2}, which allow to determine the parameter combination $HR$ as a function of the tension $\bar \lambda$,  are the main analytical result of our work.}
For small velocities $\tilde v \ll 1$ we find a linear dependence
\begin{align}\label{eq:lin_dep}
	HR \approx \bar \lambda -1 \;.
\end{align}

\begin{figure}[t]
	\centering
	\includegraphics[width=0.45\textwidth]{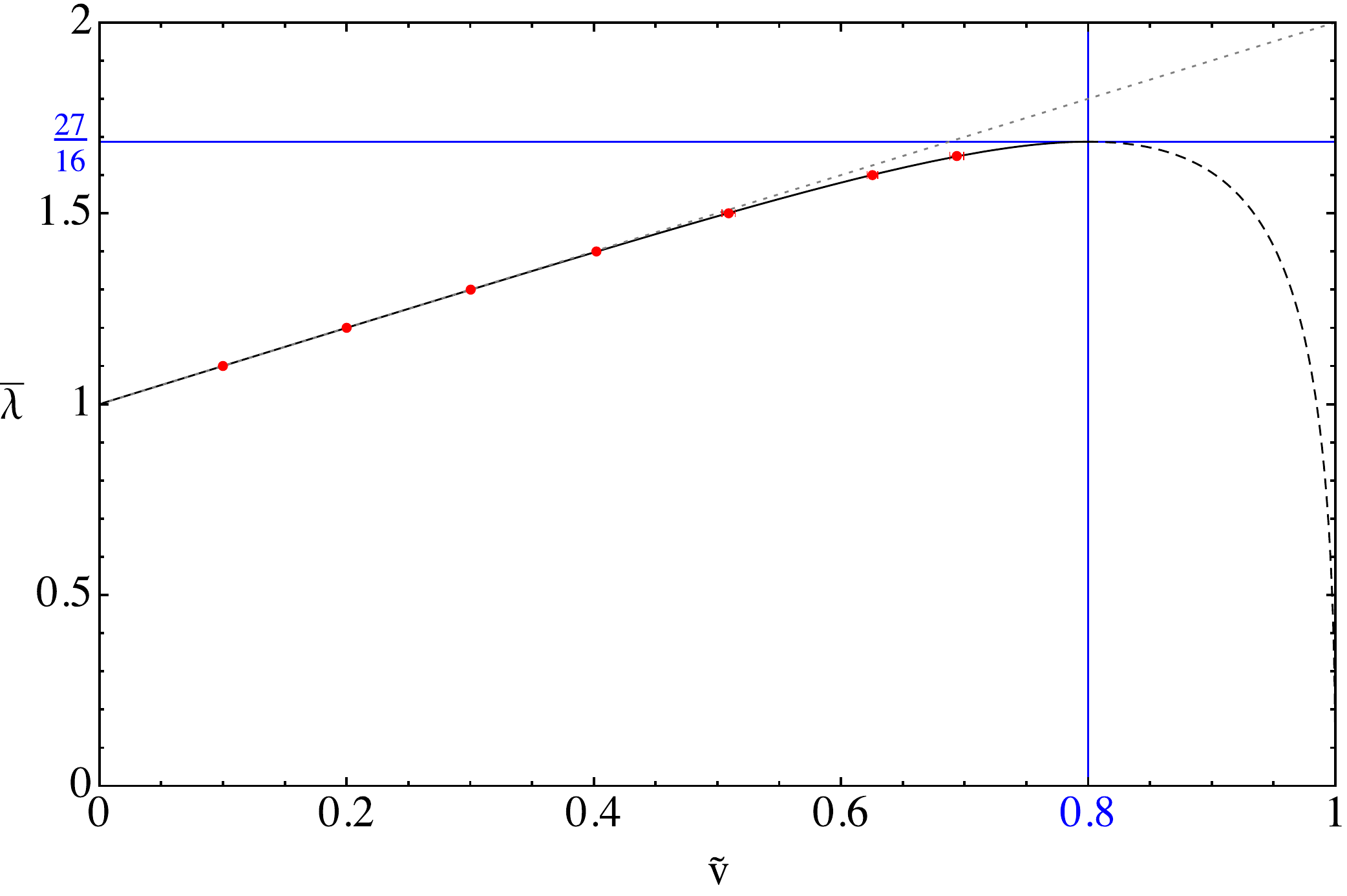}
	\caption{There are scaling solutions only for $\bar \lambda < 27/16 \approx 1.69$. Above that value the system \eqref{eq:system1} and \eqref{eq:system2} has no real solution. Below, there are two branches, one corresponding to $\tilde v < 4/5$ and the other to $\tilde v > 4/5$.  The red dots, showing the numerical results, single out the former branch as being the attractor solution and thus physically interesting.}
	\label{fig:vRhoPlot}
\end{figure}

\begin{figure}[t]
	\centering
	\includegraphics[width=0.45\textwidth]{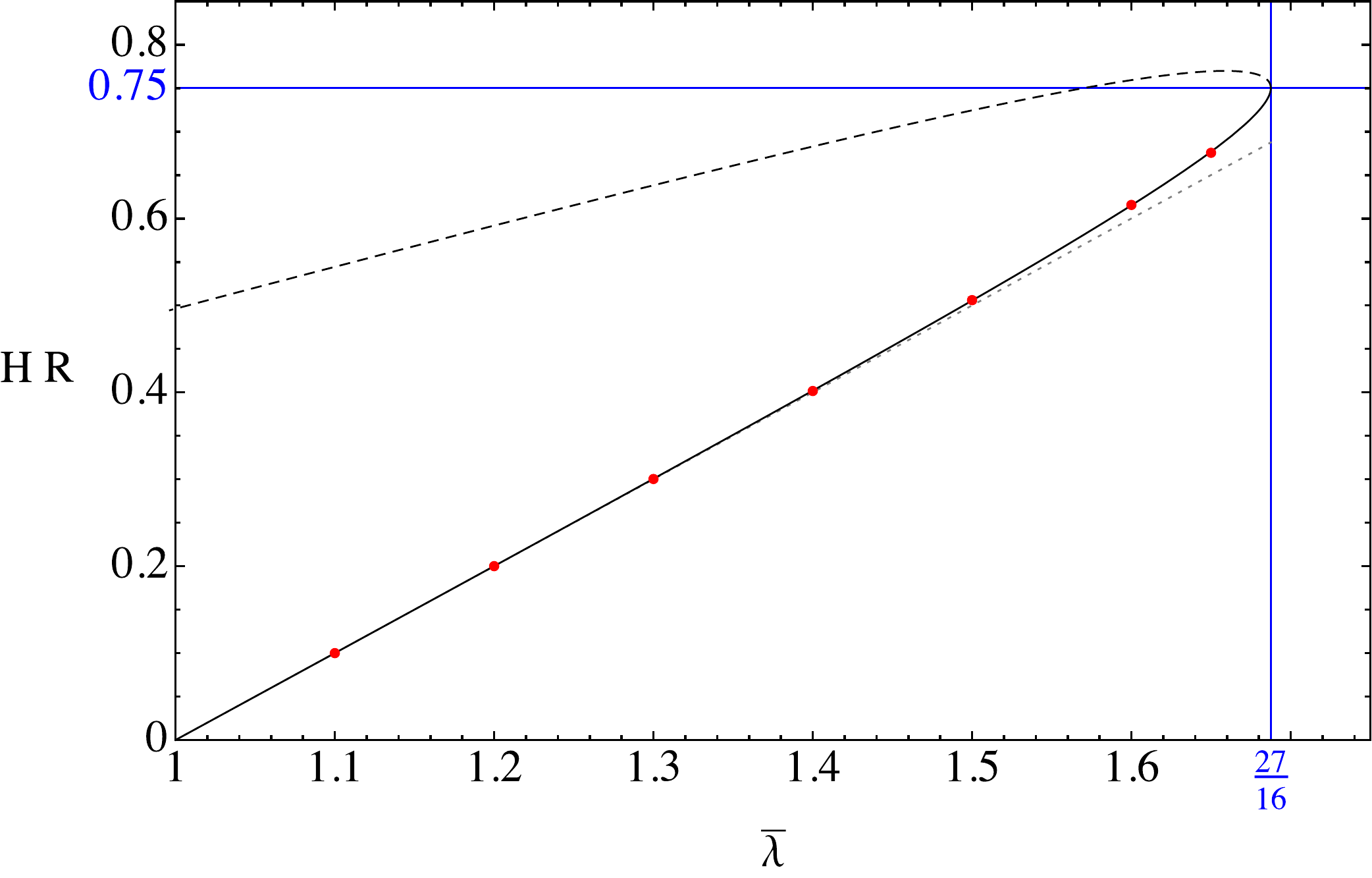}
	\caption{The axial expansion rate of the super-critical string as a function of the tension $\bar \lambda$. The linear relation \eqref{eq:lin_dep} corresponds to a good approximation. }
	\label{fig:rhoHPlot}
\end{figure}

The exact system does not possess solutions for arbitrarily large tensions. In fact, there is the rather stringent bound  $\bar \lambda<27/16 \approx 1.69 $ corresponding to a velocity $\tilde v=4/5$. Below that value there are two branches of solutions corresponding to $\tilde v<4/5$ and $\tilde v>4/5$. Only the former branch turns out to be an attractor. These results are summarized in Fig.~\ref{fig:vRhoPlot} and \ref{fig:rhoHPlot}. The solid line depicts the analytical result for the attractor branch, whereas the dashed line depicts the other branch. Each dot corresponds to one run of the numerics for different values of $\bar \lambda$. Of course, the number of time step was chosen such that the convergence of $H$ and $\tilde v$ was sufficiently accurate. The corresponding error bars, usually not exceeding the size of the dots, are also shown. The dots lie almost perfectly on the solid line which shows that we have indeed found the correct attractor solutions. Moreover, Fig.~\ref{fig:rhoHPlot} nicely illustrates that the linear dependence of Hubble on $\bar \lambda$ in \eqref{eq:lin_dep}, drawn as a dotted line, is a very good approximation for almost the whole regime. 

\begin{figure}[t]
	\centering
	\includegraphics[width=0.45\textwidth]{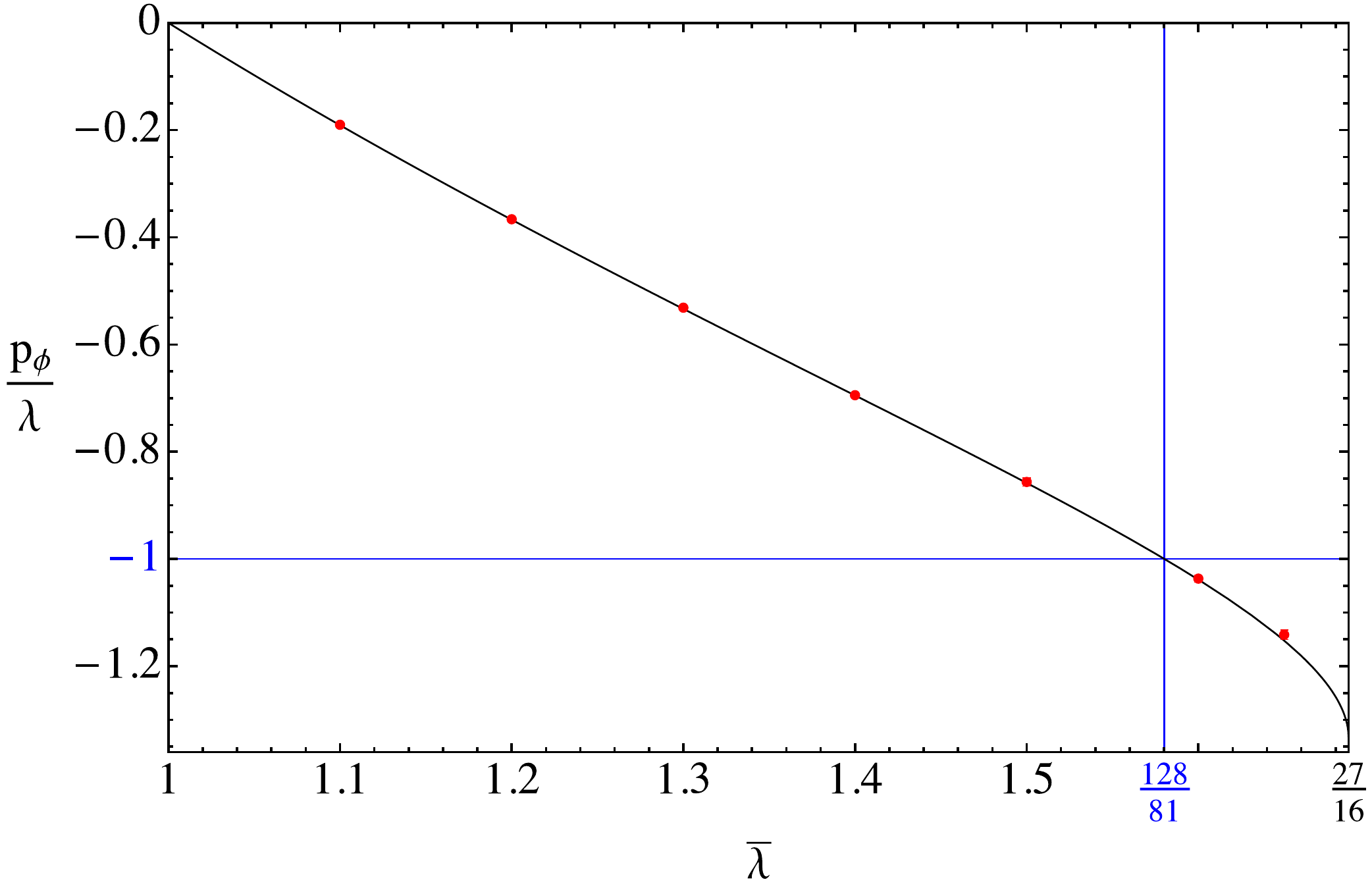}
	\caption{The string can be radially stabilized by physical matter as long as $ \bar \lambda < 128/81 \approx 1.58 $. For larger values of the tension the equation of state drops below $-1$ indicating that a stabilized shell cannot be realized physically.}
	\label{fig:rhoPPhiPlot}
\end{figure}

At first sight, the physical origin of the bound  $\bar \lambda<27/16$ is unclear because it looks as if the system cannot be solved  for larger tensions. This puzzle can be resolved by calculating the pressure in $\phi$-direction, which we implemented to stabilize the physical circumference of the shell. This can be done by evaluating \eqref{eq:phiphi_junc_cond} for the scaling solutions described above. The resulting relation between $p_{\phi}$ and $\bar \lambda$ is shown in Fig.~\ref{fig:rhoPPhiPlot}. We see that for $\bar \lambda < 128/81 \approx 1.58 $, the equation of state of $p_{\phi}$ satisfies the Null Energy Condition, i.e.\ it is greater than $-1$. This means that the shell can be stabilized by means of physically reasonable matter. However, for $\bar \lambda > 128/81 $ --- which happens \textit{before} the maximum value $ 27/16 $ is reached --- the equation of state drops below $-1$, indicating that it is no longer possible to have a stabilized shell. Consequently, we should not trust the scaling solutions in this regime because their derivation relied explicitly on that assumption. In this regime a different approach that allows for an angular expansion of the shell is needed. This would require to go beyond the effective shell description of the transverse sector and could be achieved by studying the full Nielsen-Olesen setup as done numerically in \cite{Cho:1998xy}. This work also allows for a non-trivial cross-check of our effective description. More precisely, in \cite{Cho:1998xy} it was found that a string with unit winding number begins to expand in transverse directions once\footnote{The translation to our variables is achieved by the identification $\bar \lambda=8 \pi |n| \eta^2 / m_p^2$ which holds in the Bogomol'nyi limit. } $\bar \lambda > 1.57 \pm 0.06$. This is in perfect agreement with our result. Note that at this point $ HR $ is already close to one, and we are therefore no longer insensitive to the microscopic details of the string. The perfect numerical agreement between both approaches thus seems to be an  accident, and indeed, for higher winding numbers, the stability bound derived in the NO-setup is slightly below the one derived in our setup. This demonstrates that quantitative predictions get sensitive to the underlying UV model once $ HR $ is of order one.

We also checked explicitly that the numerically determined radial profiles of $\tilde \alpha$ and $\tilde \eta$ approach the analytic solutions \eqref{eq:alpha_scal} and \eqref{eq:eta_scal}, respectively. Within the numerical error bars, we found perfect agreement after the system was evolved sufficiently far in time.

Another consistency check of our analysis concerns the cross-sectional area of the interior space. As has been argued before, the area should be constant for the regularization to work properly. After substituting the scaling solution in \eqref{eq:area}, we find
\begin{align}
\tilde A=\pi R^2\frac{ \left(1+\sqrt{1-\tilde v^2}\right)^2}{4 \tilde v^2}\ln\left(1-\tilde v^2 \right)\;,
\end{align}
which reduces to the flat space result in the limit $\tilde v \rightarrow 0$ as expected. Most importantly, this expression is time independent (and finite for $ \tilde v < 1 $) showing that our regularization scheme is stable. The analytical value is depicted in Figure \ref{fig:volIn_supercrit} as a dashed line. It is approached by the numerical solution consistently.

The fact that the area inside the shell is exactly constant for the scaling solution suggests that the radial movement of the shell is just a coordinate relict. Accordingly, we will show in the Appendix \ref{ap:cigar_coord} that we can introduce new coordinates for which the shell sits at a constant coordinate position and the whole time dependence of the metric is related to the expansion in axial direction. Moreover, we will show that the scaling solution is equivalent to a solution discussed by Witten~\cite{Witten:1982} and Gregory~\cite{Gregory:2003xf}, describing a ``cigar'' shaped universe. In both works the vacuum solution was discussed but without matching it to an actual matter model. To our knowledge this has been achieved for the first time within our super-critical string setup. 

One might wonder whether the analytic scaling solution discussed above could also describe the asymptotic form of the exterior spacetime by simply replacing $ \tilde v $ by $ v = 1 $. However, this cannot work because eq.~\eqref{eq:system2} could then also be derived from the exterior\footnote{It would only differ by the finite and constant overall factor $ \exp(-\eta|_{r=0}) $ corresponding to the conical singularity at the exterior axis.} and would thus give the contradictory result $ HR = 0 $. Thus, the actual exterior solution cannot converge to the $ v = 1 $ scaling solution, at least not everywhere. This can also be understood by realizing that in the scaling solution the shell moves at constant (coordinate) velocity; hence, for $ v = 1 $ it would always move exactly with the speed of light, which is impossible for a massive shell (and would also contradict $ \tilde v < 1 $, as already mentioned earlier). In other words, the actual attractor solution in the exterior would have to be one in which the shell's speed is not constant but only approaches 1 at late times, and can thus not be the simple scaling solution.

Nevertheless, we found that the numerical solution does indeed approach the $ v=1 $ scaling solution for most $ r $ including the axis, but starts to deviate from it close to the shell. Moreover, the concrete form of these deviations remains sensitive to the initial conditions for all times, so it seems impossible to make any further generic statements about the full exterior attractor solution.

The convergence towards the scaling solution sufficiently far away from the string can also be seen qualitatively in the embedding diagrams in Fig.~\ref{fig:embed_supercrit} and \ref{fig:embed_supercrit_reg}, which nicely agree with the cigar shape of the scaling solution. The cigar keeps growing and presumably becomes infinitely long as $ \tau \to \infty $, cf.\ Fig.~\ref{fig:volOut_supercrit} and \ref{fig:volOut_supercrit_reg}.

Finally, let us emphasize that the relation we derived here relates the expansion rate to the tension of the (regularized) super-critical string. The conical singularity at the exterior axis can be interpreted as another (unregularized) sub-critical string. This point of view was for instance taken in ref.~\cite{Gregory:2003xf}. However, the corresponding deficit angle --- and thus the tension of this sub-critical string --- is not generically related to the expansion rate. Instead, it can be chosen independently, and in particular even be set to zero, as we will show in the next section (see also Fig.~\ref{fig:embed_supercrit_reg}). In summary, \textit{a sub-critical string only creates a deficit angle and does not inflate, whereas a super-critical string inflates in axial direction at the rate determined by the system \eqref{eq:system1} and \eqref{eq:system2}.}

Our results are also relevant for 6D braneworld models. In this case the string is replaced by a 3-brane corresponding to our universe. A generalized version of equation \eqref{eq:lin_dep} then plays the role of a modified Friedmann equation. This higher dimensional picture is discussed in Section~\ref{sec:6D}.
\section{Removing the conical singularity} \label{sec:remove_conical_sing}

In Sec.~\ref{sec:static}, we saw that for super-critical string tensions, the static solution necessarily has a conical singularity at the exterior axis. This means that the vacuum Einstein equations are actually not satisfied there; instead, there is a second (unregularized, sub-critical) string sitting at the axis, the tension of which must be suitably dialed according to the tension of the original (regularized, super-critical) string. This is of course rather unsatisfactory, because we wanted to model a single super-critical cosmic string. Physically, there is no reason why the second string should be necessary.

This suggests that the second string --- or equivalently, exterior conical singularity --- is an artifact caused by the too strong assumption of having a static geometry. And indeed, for time dependent solutions, this need not be the case. The absence of a conical singularity is equivalent to $ \eta|_{r=0} = 0 $. Using the vacuum field equation \eqref{eq:ER_vac_eta_r}, we can rewrite this as
\begin{equation} \label{eq:no_conical_sing}
	\eta|_{r=0} = \eta_0 - \int_0^{r_0} r \left[ \left(\partial_t\alpha\right)^2 + \left(\partial_r\alpha\right)^2 \right] \rd r \stackrel{!}{=} 0 \,.
\end{equation}
For (regular) static solutions, $ \alpha $ is constant and so $ \eta|_{r=0} = \eta_0 $, which is non-zero. But for time dependent solutions, the integral in \eqref{eq:no_conical_sing} is positive. Hence, if $ \eta_0 > 0 $, one can always chose initial conditions for $ \alpha $, such that~\eqref{eq:no_conical_sing} is fulfilled at the initial time. But then it will in fact be fulfilled for all times, since the constraint~\eqref{eq:ER_vac_eta_t} implies that $ \partial_t\eta|_{r=0} = 0 $ (if $ \alpha $ is regular, which we assume). Whether $ \eta_0 $ is positive, again depends on the string tension and the initial conditions for $ \tilde\alpha $. Specifically, using the junction condition \eqref{eq:00_junc_cond}, one can show that (for super-critical tensions) $ \eta_0 > 0 $ is equivalent to
\begin{equation} \label{eq:conic_sing_bound}
	\bar\lambda < \tilde \gamma + \sqrt{1 + H^2 R^2} \,.
\end{equation}
Thus, if $ \bar\lambda $ lies inside the non-empty interval
\begin{equation} \label{eq:regular_region}
	\tilde\gamma + |H| R < \bar\lambda < \tilde\gamma + \sqrt{1 + H^2 R^2} \,,
\end{equation}
the conical singularity at the exterior axis can always be removed by a suitable choice of initial data for $ \alpha $.

In the numerical examples that we studied, condition~\eqref{eq:conic_sing_bound} was always satisfied. Indeed, for the flat profile function $ F(x) = 1 $, equation \eqref{eq:eta0Tilde_init} implies that the bound~\eqref{eq:conic_sing_bound} is well above the bound $ \bar\lambda < 27/16 $, beyond which the solutions discussed above are no attractors anyway, see Fig.~\ref{fig:regionPlot}. 

However, the rather arbitrary choice of initial profiles~\eqref{eq:init_profiles_supercrit} does not automatically lead to a smooth axis. But we checked that the late time asymptotic behavior, as well as the relation between string tension and expansion rate, still persist if the initial data for $ \alpha $ is deformed  such that \eqref{eq:no_conical_sing} is satisfied and the exterior geometry is perfectly smooth.

In other words, the deficit angle at the exterior axis corresponds to a static, sub-critical string that is put there in addition to the actual super-critical string of interest. Its tension is a parameter that is completely controlled by the initial data; in particular, it can be set to zero whenever \eqref{eq:regular_region} is satisfied. Furthermore, the expansion rate $ H $ is completely insensitive to this sub-critical string and is instead set by the tension of the super-critical string. This can also be understood from the formation of the horizon: an observer co-moving with the cylindrical shell cannot even see the axis, and hence cannot tell whether there is a conical singularity or not. Therefore, intrinsic quantities like $ H $ cannot depend on the exterior defect angle either.

The fact that the conical singularity can be removed in the dynamical case, while being physically satisfactory, raises another question: How does the system know which side of the shell is the interior and which the exterior? After all, both regions are described by the same metric ansatz and share the same boundary conditions at the axes. Still, the numerical results show that the shell's velocity approaches unity only in the exterior, implying that the two regions do in fact evolve differently. Clearly, this difference must already be incorporated in the initial conditions. If those were completely symmetric as well, no difference between ``inside'' and ``outside'' could ever emerge.

And indeed there is such a difference: the difference between $ \tilde\eta_{0i} $ and $ \eta_{0i} $, measuring the gravitational C-energy in the interior and exterior, respectively. In the case in which the exterior conical defect was removed, this difference was caused by choosing a non-trivial initial profile for $ \alpha $ only in the exterior. In the case with conical singularity, both initial profiles were chosen identically, but the localized energy-density corresponding to the conical defect also adds to the exterior C-energy, while there is no such contribution in the interior. Thus, in both cases $ \eta_{0i} $ was larger than $ \tilde\eta_{0i} $, or in other words, there was more C-energy in the exterior than in the interior.

The symmetry of the setup then implies that the opposite also holds: If we interchange initial conditions, such that there is more C-energy in the ``interior'', then the velocity will approach $ 1 $ there, and a constant $ < 1 $ in the ``exterior''. But in this situation, we would simply interchange the names ``interior'' and ``exterior'', because there should not be a horizon in the interior if we want to view it as a regularization of a thin cosmic string. Hence, the initial conditions should always be chosen such that there is less C-energy in the interior; or equivalently, the side with less C-energy should be identified as the actual interior. Note that, in particular, this qualification of inside and outside does not depend on whether there is a conical singularity at either axis.

As already remarked, in the completely symmetric case $ \tilde\eta_{0i} = \eta_{0i} $ there can be no difference between inside and outside. In fact, we found that there is a finite region\footnote{We did not investigate this more quantitatively.} $ \tilde\eta_{0i} \approx \eta_{0i} $ for which no difference emerges. In these cases \textit{both} velocities approach unity, and so one cannot identify an interior, and can thus not speak of a regularized cosmic string. We therefore did not further investigate these cases.

\section{Parameter plot}\label{sec:parameter_plot}

\begin{figure}[t]
	\centering
	\includegraphics[width=0.45\textwidth]{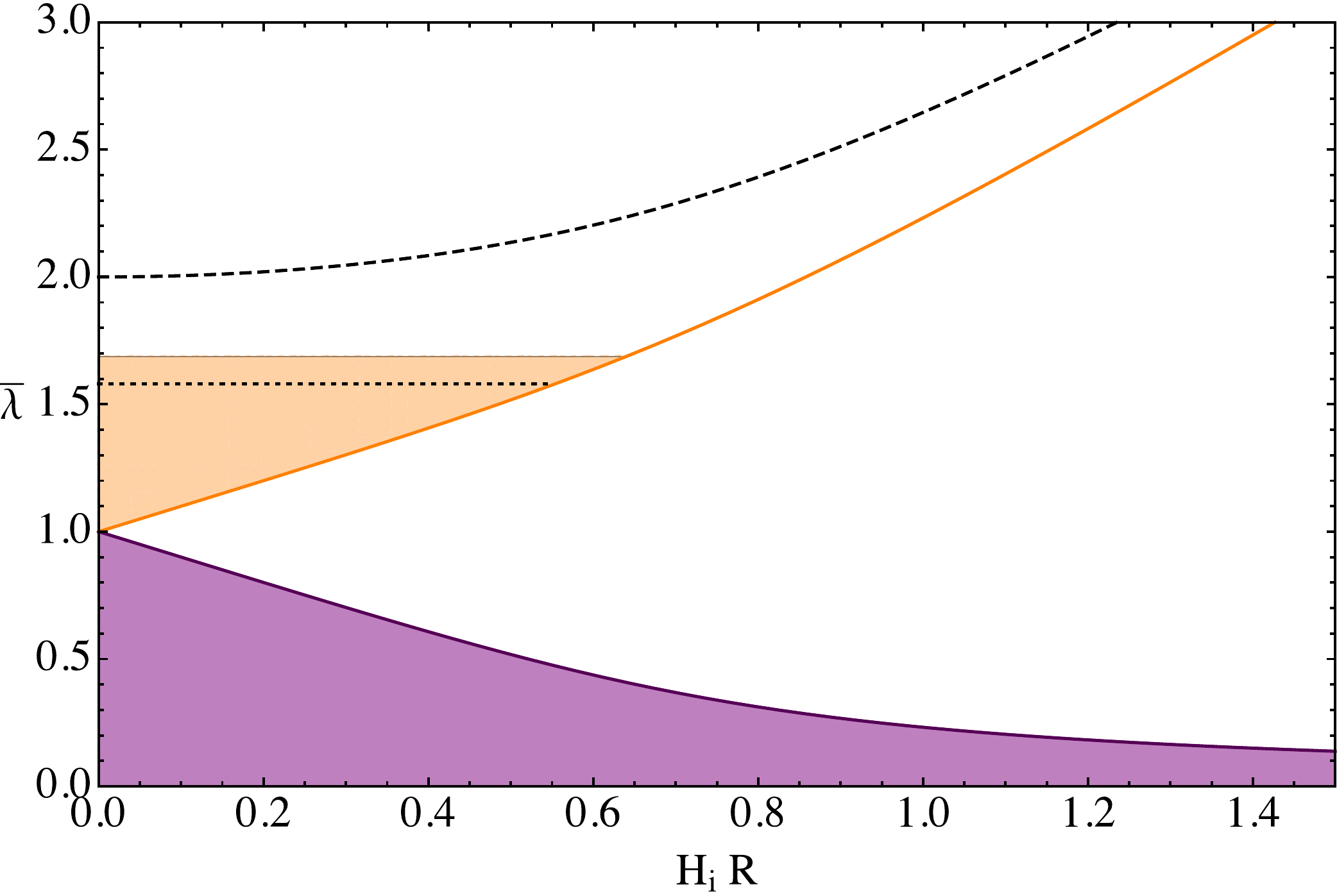}
	\caption{(Color online) Region plot of parameter space, as explained in the text.}
	\label{fig:regionPlot}
\end{figure}

Our findings can nicely be summarized in the parameter plot shown in Fig.~\ref{fig:regionPlot}.
It assumes a trivial radial profile of $ \tilde\alpha $, which we chose as initial data for the numerics. (If the radial profile is non-trivial, $ \tilde\eta_0 $ is not rigidly related to $ HR $, and so the parameter space becomes three-dimensional. Hence, this plot is only valid at initial time.)

In the purple (dark gray) region, the string is sub-critical and the static defect angle solution is an attractor. Between the solid purple (dark gray) and orange (light gray) lines, the system is critical; this region is not covered in the present work. Above the orange line, the string is super-critical, and in the shaded region it approaches the axially expanding scaling solutions discussed in Sec.~\ref{sec:num_supercrit} and~\ref{sec:analytic}. The upper bound of this region is $ \bar\lambda = 27/16 \approx 1.69 $, beyond which there is no solution for $ \tilde v(\bar\lambda) $ anymore and hence the geometry cannot approach the scaling solutions, cf.\ Sec.~\ref{sec:analytic}. However, before this bound is reached, the equation of state for $ p_\phi $ becomes less than $ -1 $ for $ \bar\lambda = 128/81 \approx 1.58 $ (dotted black line), which should be interpreted as the statement that the string thickness cannot be stabilized anymore, in accordance with \cite{Cho:1998xy}. Therefore, we did not further investigate the upper white region. Finally, the dashed black line corresponds to the bound~\eqref{eq:conic_sing_bound}, beyond which the conical singularity at the exterior axis could not be removed. It is well above the orange region, implying that in the cases we studied, which dynamically approach the analytic scaling solutions of Sec.~\ref{sec:analytic}, the conical singularity can always be removed.

\section{Braneworld in 6D} \label{sec:6D}
Our results are in particular interesting with respect to six dimensional braneworld scenarios where the string is promoted to a 3-brane and the coordinates $(r,\phi)$ label the two extra-dimensions. Related models have been studied extensively in the past for both compact~\cite{Burgess:2011va} as well as infinite extra dimensions~\cite{Dvali:2002pe}; for a more recent work see~\cite{Niedermann:2014bqa}. The general hope is to find solutions for which the gravitational impact of a brane tension or, equivalently, a 4D cosmological constant, is fully absorbed by the extrinsic curvature corresponding to the embedding of the brane in the bulk. Such a mechanism would make the cosmological constant invisible to a brane observer and would therefore allow to address the cosmological constant problem~\cite{Weinberg:1988cp}. 

In this section, we present the corresponding super-critical solutions. The Einstein-Rosen coordinates~\eqref{eq:met_ER} can be easily generalized to six dimensions by making the formal replacements $\alpha \rightarrow 3\alpha$ in the first and last term and $z \rightarrow \vec{x} $ in the middle term of \eqref{eq:met_ER}. Here $ \vec{x} $ denotes the three spatial brane directions. In the exterior region, for instance, we get
\begin{align}
	\rd s^2 & = \re^{2(\eta - 3\alpha)} \left( -\rd t^2 + \rd r^2 \right) + \re^{2\alpha} \rd \vec{x}^2 + \re^{-6\alpha} r^2 \rd\phi^2 \,.
\end{align}
The corresponding vacuum equations and junction conditions can be derived in complete analogy to the 4D case presented above, and the system can then be solved numerically in a similar fashion. The results that we found are also analogous:
The axial expansion rate $H$, which in this case corresponds to the ordinary Hubble parameter, approaches again a constant, non-zero value. Furthermore, the coordinate velocity of the brane exhibits the same asymptotic behavior \eqref{eq:limit_attract} as in the 4D case, thus implying
\begin{align} \label{eq:gamma_rel_6D}
	\tilde\gamma \to \frac{3 H R}{\tilde v} \,, && \gamma \to 3 H R \,. 
\end{align}
Using the generalized version of \eqref{eq:00_junc_cond}, we find
\begin{align}\label{eq:system1_6D}
	\bar \lambda = 3 H R \left(1+\frac{1}{\tilde v} \right)\;,
\end{align}
where now $ \bar\lambda := \lambda / \left( 2\pi M_6^4 \right) $, with $ M_6 $ denoting the six-dimensional Planck mass and $ \lambda $ the energy per 3D string volume, corresponding to a 4D cosmological constant or vacuum energy.
The analytic solutions in the interior can be derived in the same way as before. We find that $\tilde \alpha$ is still of the form \eqref{eq:alpha_scal} where the same expression \eqref{eq:alpha_scaling} for $\Omega$ holds.
For $\tilde \eta$ we find
\begin{align}
	\tilde \eta(\tilde t, \tilde r) = \frac{4}{3} \ln\left(\frac{ 1+\sqrt{1-x^2} }{2\sqrt{1-x^2}}\right)\;,
\end{align}
which enables us to generalize \eqref{eq:system2} to
\begin{align}\label{eq:system2_6D}
	3HR =  \tilde v \left(1-\tilde v^2\right)^{\frac{1}{6}}\left(\frac{2}{1+\sqrt{1-\tilde v^2}}\right)^{\frac{4}{3}}\;.
\end{align}
As before, in the physically relevant regime where $HR \ll 1$, we can analytically eliminate $\tilde v$ from the system \eqref{eq:system1_6D} and \eqref{eq:system2_6D}, yielding a modified Friedmann equation
\begin{align} \label{eq:modFried}
	3 HR \approx \bar \lambda -1 \;.
\end{align}
It should be noted that for a realistic value of the regularization scale, say $R \sim {\rm TeV^{-1}}$, this equation requires again a tremendous amount of fine tuning between the two terms on the right hand side in order to describe the observed accelerated expansion of the universe. Thus, these super-critical solutions within pure six dimensional GR cannot help with the cosmological constant problem. The phenomenology of this equation in the context of a more sophisticated braneworld model~\cite{Dvali:2002pe,Niedermann:2014bqa} will be discussed elsewhere.

\section{Conclusion and outlook}\label{sec:conclusion}
In this work, we have studied the geometry of a single super-critical cosmic string. In a marginally super-critical regime, $1<\bar \lambda \lesssim1.6$, the string can consistently be modeled as a cylindrical shell of fixed circumference $2 \pi R$. Within this parameter region, there are well known static solutions for which the geometry is compact and closes in a second singular axis away from the string. By numerically solving the full system of vacuum and shell matching equations, the instability of the static solution was demonstrated. It was shown that the system instead approaches a time-dependent attractor solution with the following properties:
\begin{itemize}
\item The string expands in axial direction at a constant rate.

\item A horizon is formed away from the string.

\item The bulk geometry remains compact but becomes cigar-shaped and expands. 

\item In the dynamical solutions, the exterior conical singularity can always be avoided.
\end{itemize}

Moreover, an analytic relation between the tension $\lambda$, the string thickness $R$ and the expansion rate $H$ was derived. 

Generalized to six dimensional GR, these solutions correspond to a simple braneworld model with a single super-critical, pure tension brane. The corresponding modified Friedmann equation is given by Eq.~\eqref{eq:modFried}. Unfortunately, the parameters of the model have to be fine-tuned to be in accordance with the observed accelerated expansion of the universe. As a future direction, it would be interesting to extend these solutions to different more sophisticated braneworld models. In particular, the analysis in \cite{Kaloper:2007ap, Niedermann:2014bqa} in the context of ``Brane Induced Gravity'' (BIG) models in six dimensions can now be extended to super-critical brane tensions. As another interesting application, our solutions allow to directly generalize the 5D model in~\cite{Kaloper:2007my}, describing an effective Kaluza-Klein setup at small length scales, to the super-critical case. Furthermore, we are planning to generalize our results to BIG-branes with more than two co-dimensions. Those setups are discussed in~\cite{Dvali:2002pe} as potential solutions to the cosmological constant problem. Ultimately, this will enable us to clarify their phenomenological status.

\begin{acknowledgments}
We thank Tony Gherghetta, Gary W.\ Gibbons, Stefan Hofmann, Nemanja Kaloper, Justin Khoury and Marco Peloso for helpful discussions.
The work of FN was supported by the DFG cluster of excellence `Origin and Structure of the Universe' and  by TRR 33 `The Dark Universe'.
The work of RS was supported by the DFG cluster of excellence `Origin and Structure of the Universe'.
\end{acknowledgments}

\appendix

\section{Classification of exterior geometries} \label{ap:ext_geom}

In this appendix, we give a complete classification of the character of a vacuum spacetime described by the metric \eqref{eq:met_general}, depending on the gradient $ \nabla W \equiv (\partial_{t^*}W, \partial_{r^*}W) $. Furthermore, we discuss which changes of character are admissible in vacuum, as well as across shells of matter and show how the change of character across such a shell depends on the surface energy density.

\begin{table}[ht]
\centering
\begin{tabular}{ | l | l | l | l | }
\hline
\textbf{Character} & \textbf{Orientation of $ \nabla W $} & \textbf{$ W_+' $} & \textbf{$ W_-' $} \\
\hline \hline
$ D^{+} $ & space-like outward & $ > 0 $ & $ < 0 $ \\
\hline
$ D^{-} $ & space-like inward & $ < 0 $ & $ > 0 $ \\
\hline
$ D^{\uparrow} $ & time-like future & $ > 0 $ & $ > 0 $ \\
\hline
$ D^{\downarrow} $ & time-like past & $ < 0 $ & $ < 0 $ \\
\hline
$ D^{+\uparrow} $ & light-like outward-future & $ > 0 $ & $ = 0 $ \\
\hline
$ D^{+\downarrow} $ & light-like outward-past & $ = 0 $ & $ < 0 $ \\
\hline
$ D^{-\uparrow} $ & light-like inward-future & $ = 0 $ & $ > 0 $ \\
\hline
$ D^{-\downarrow} $ & light-like inward-past & $ < 0 $ & $ = 0 $ \\
\hline
$ D^{\times} $ & zero & $ = 0 $ & $ = 0 $ \\
\hline
\end{tabular}
\caption{Definition of spacetime character, depending on the gradient of $ W $.} \label{tab:gradW}
\end{table}

Following Thorne~\citep[Appendix A]{PhysRevD.46.2435}, we define the ``character'' of spacetime depending on the orientation of $ \nabla W $ as summarized in Table~\ref{tab:gradW}. Here and henceforth, ``outward'' (resp. ``inward'') means in direction of increasing (decreasing) $ r^* $, and ``future'' (``past'') refers to the direction of increasing (decreasing) $ t^* $. In vacuum, $ W $ satisfies the 1D wave equation~\eqref{eq:1D_wave}, the general solution of which can be written as
\begin{equation} \label{eq:W_sol}
	W(t^*, r^*) = W_+(t^* + r^*) + W_-(t^* - r^*) \,.
\end{equation}
It easy to verify that each orientation of $ \nabla W $ corresponds to a certain choice of signs for the derivatives\footnote{Here and henceforth, the primes acting on $ W_{\pm} $ denote the derivative with respect to their argument (not with respect to $ r^* $), i.e.\ $ W_{\pm}'(x) := \rd W_{\pm}(x) / \rd x $.} $ W_+' $ and $ W_-' $, as listed in Table~\ref{tab:gradW}.

\subsection{Changes of character in vacuum}

Let us now discuss which changes of character are allowed in vacuum regions. The above discussion shows that a change of character is equivalent to a change of sign of $ W_+' $ or $ W_-' $. Thus, on the boundary between two spacetime regions of different character, we have $ W_+' = 0 $ or $ W_-' = 0 $. But since these functions are constant along null-surfaces, it follows that \textit{in vacuum, the character can only change across null-surfaces}.

\begin{figure}[t]
	\centering
	\includegraphics[width=0.35\textwidth]{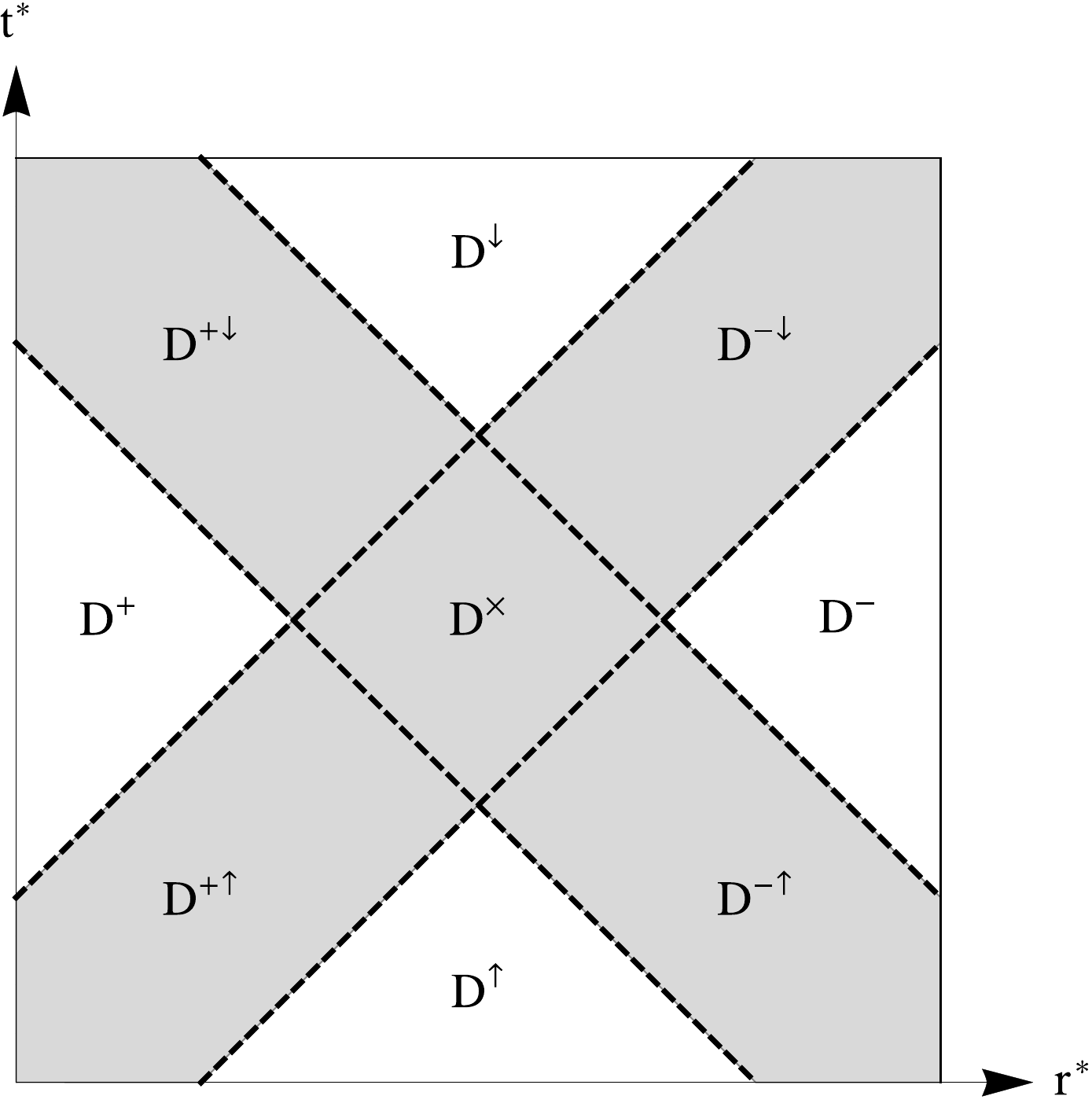}
	\caption{All admissible changes of character in vacuum regions can be read off from this diagram, by following incoming or outgoing null rays.}
	\label{fig:character_changes}
\end{figure}

Furthermore, since $ W_+' $ is constant along an incoming light-ray ($ \rd t + \rd r = 0 $), the only possible changes along incoming rays are those in which $ W_-' $ changes sign, i.e.\ changes within the following sets:
\begin{align*}
	\{ D^+, D^\uparrow, D^{+\uparrow} \}, && \{ D^-, D^\downarrow, D^{-\downarrow} \}, && \{ D^{+\downarrow}, D^{-\uparrow}, D^\times \} \,.
\end{align*}
Similarly, along outgoing null-rays ($ \rd t - \rd r = 0 $), the character can only change within
\begin{align*}
	\{ D^+, D^\downarrow, D^{+\downarrow} \}, && \{ D^-, D^\uparrow, D^{-\uparrow} \}, && \{ D^{+\uparrow}, D^{-\downarrow}, D^\times \} \,.
\end{align*}
However, there is a further restriction on the directions of changes within all these sets, coming from the focusing theorem for null geodesics~\cite{PhysRevD.46.2435, Misner}: It says that $ \rd^2 W / \rd\sigma^2 \leq 0 $, where $ \sigma $ is the affine parameter along the null geodesic. Hence, the functions $ W_+' $ and $ W_-' $ cannot get larger along any null rays. Thus, e.g.\ for the first set, the only admissible changes are
\begin{equation}
	D^\uparrow \to D^{+\uparrow} \to D^+\,,
\end{equation}
and similarly for the other sets. All changes that are finally possible, are summarized schematically in Fig.~\ref{fig:character_changes}: they are exactly those changes which are encountered along any incoming or outgoing null ray in this diagram. Note that in the gray shaded regions the Jacobian of the transformation \eqref{eq:ER_gauge} vanishes, and so these coordinates can only be adopted in any of thef white regions separately. Furthermore, the new coordinate $ r $ which is set equal to $ W $ will be a spatial coordinate for $ D^+ $ and $ D^- $, but a temporal coordinate for $ D^\uparrow $ and $ D^\downarrow $. In a $ D^+ $ (resp.\ $ D^- $) region, $ r $ decreases as one moves inward (outward); it can not start increasing again, because this would require a change to $ D^- $ ($ D^+ $), which is not admissible. Hence, it decreases until eventually $ r=0 $, implying an axis that delimits spacetime, as advertised in Sec.~\ref{sec:supercrit}. (Similar statements hold in the temporal case, but there the ``axes'' correspond to physical, initial and final collapse singularities.)

\begin{figure}[t]
	\centering
	\includegraphics[width=0.35\textwidth]{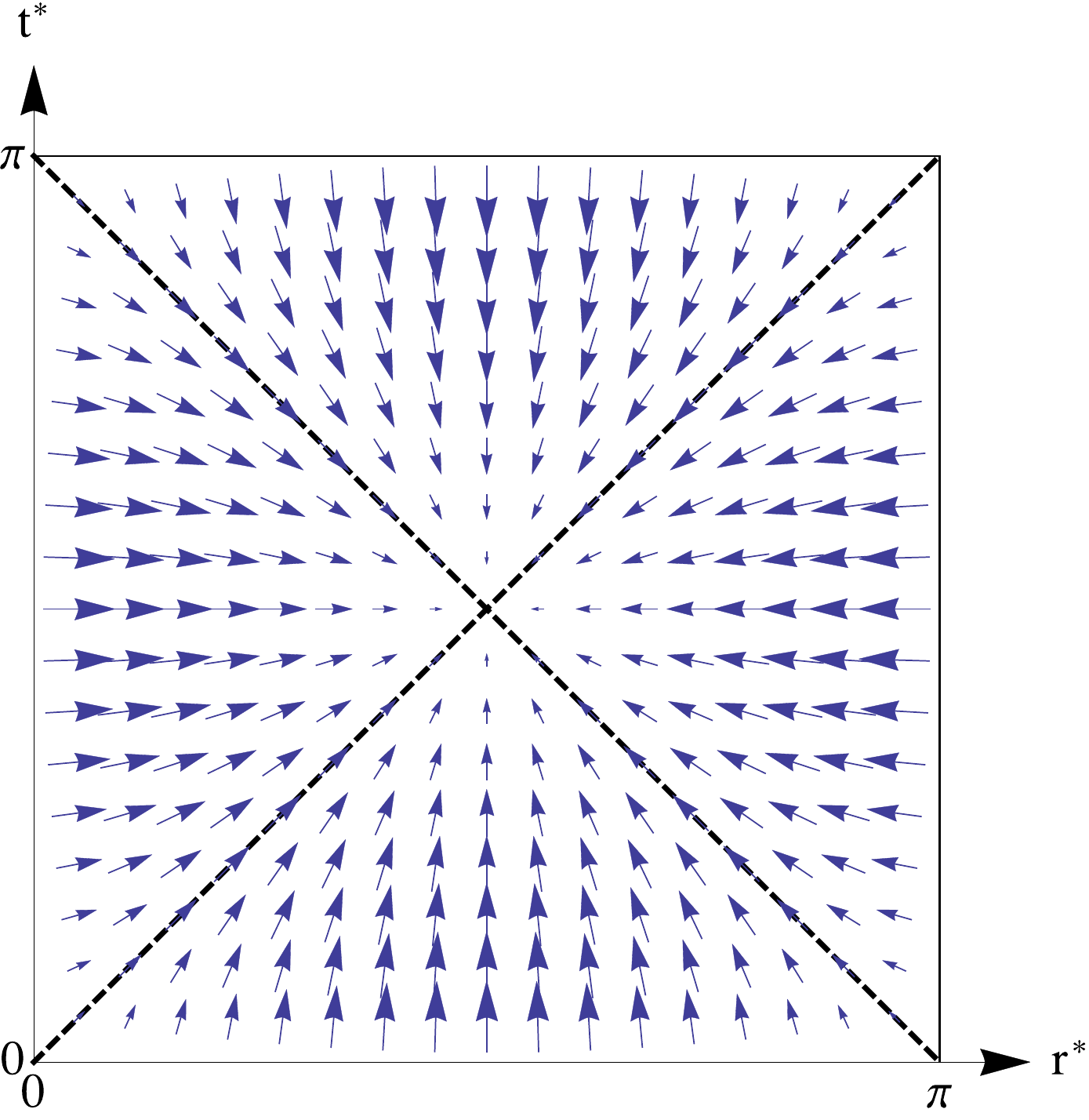}
	\caption{The gradient of $ W $ for the Gowdy universe, $ W = \sin(t^*) \sin(r^*) $. The corresponding spacetime character matches that of Fig.~\ref{fig:character_changes}, with the gray regions degenerated to single lines.}
	\label{fig:gowdy}
\end{figure}

There is actually a well-known example of a (vacuum) spacetime exhibiting all the different characters: the Gowdy universe~\cite{PhysRevLett.27.826}, in which $ W = \sin(t^*) \sin(r^*) $. The corresponding vector plot of $ \nabla W $ is shown in Fig.~\ref{fig:gowdy}. The spacetime character exactly matches that of Fig.~\ref{fig:character_changes}, with the gray regions degenerated to single lines. The two axes are located at $ r^* \in \{ 0, \pi \} $, whereas $ t^* \in \{ 0, \pi \} $ correspond to the Big Bang / Big Crunch singularities\footnote{By extending the range of $ t^* $ to $ (0, 2\pi) $, one could construct an example where forbidden changes like $ D^\downarrow \to D^\uparrow $ were apparently allowed. However, they would be separated by the singularity at $ t^* = \pi $, so they should be regarded as unphysical. More generally, the above argument using null geodesics implicitly assumes that no singularities are present.}.

Coming back to our general discussion, we find another important result: \textit{It is not possible for a vacuum region to dynamically evolve from $ D^+ $ to $ D^- $ or vice versa}. Therefore, a sub-critical cosmic string (for which the exterior geometry is $ D^+ $, see below) can never evolve to a super-critical string ($ D^- $ exterior). In particular, super-critical strings cannot be formed by cylindrical collapse within classical GR.


\subsection{Changes of character across the shell}

Next, let us derive the relations given in Sec.~\ref{sec:supercrit}, relating the spacetime character of the region outside the regularized cosmic string to its tension. In the interior, we assume the character to be $ D^+ $, in order to have a well-defined regularization. Thus, we can safely adopt the Einstein-Rosen coordinates~\eqref{eq:met_ER_in} there. In the exterior, however, we keep the coordinates~\eqref{eq:met_general} because there we do not know the character yet. In these coordinates, the $ (0,0) $-component of the junction conditions~\eqref{eq:israel} reads
\begin{equation} \label{eq:00_junc_cond_W}
	\bar\lambda = \tilde\gamma - \gamma^* \left( \dot r_0^* \partial_{t^*} W + \partial_{r^*} W \right)|_0 \,.
\end{equation}
Here, $ \gamma^* $ is defined similarly to $ \gamma $ in~\eqref{eq:defGamma}, i.e.\ $ \gamma^* := \exp(-\eta^*_0) / \sqrt{1 - \dot r_0^{*2}} $ and $ \dot r_0^* := \rd r_0^* / \rd t^* $. Furthermore, continuity of the metric implies $ W_0 = \tilde r_0 $, which after differentiating with respect to $ \tau $ yields
\begin{equation} \label{eq:cont_met_W}
	0 = \tilde\gamma \dot{\tilde r}_0 - \gamma^* \left( \partial_{t^*}W + \dot r_0^* \partial_{r^*}W \right)|_0 \,.
\end{equation}
Plugging the general solution~\eqref{eq:W_sol} of $ W $ into~\eqref{eq:00_junc_cond_W} and~\eqref{eq:cont_met_W}, and solving for $ W_\pm'|_0 $, we find
\begin{subequations}
\begin{align}
	W_+'|_0 & = \frac{1}{2\gamma^* \left(1 + \dot r_0^* \right)} \left( \tilde{\gamma} + H R - \bar{\lambda} \right) \,, \\
	W_-'|_0 & = \frac{1}{2\gamma^* \left(1 - \dot r_0^* \right)} \left( \bar{\lambda} - \tilde{\gamma} + H R \right) \,.
\end{align}
\end{subequations}
Note that here we used the relation \eqref{eq:r0Dot_rel} to eliminate $ \dot{\tilde r}_0 $. Since the prefactors on the right hand side are manifestly positive, inspection of Table~\ref{tab:gradW} immediately reveals that the spacetime character at the exterior boundary of the shell is:
\begin{subequations} \label{eq:D_bounds}
\begin{align}
	D^+ \quad & \Leftrightarrow \quad \bar{\lambda} < \tilde{\gamma} - \left|H\right| R \,,\\
	D^\uparrow \;\text{or}\; D^\downarrow \quad & \Leftrightarrow \quad \tilde{\gamma} - \left|H\right|R < \bar{\lambda} < \tilde{\gamma} + \left|H\right| R \,, \label{eq:cond_timelike}\\
	D^- \quad & \Leftrightarrow \quad \bar{\lambda} > \tilde{\gamma} + \left|H\right| R \,,
\end{align}
\end{subequations}
thus verifying the result stated in Sec.~\ref{sec:supercrit}.
The orientation in the time-like case \eqref{eq:cond_timelike} depends on the sign of $ H $: it is $ D^\uparrow $ for $ H > 0 $ and $ D^\downarrow $ for $ H < 0 $. The light-like cases correspond to the saturation of one of the inequalities, e.g.\ $ D^{+\uparrow} \Leftrightarrow \bar{\lambda} = \tilde{\gamma} - \left|H\right| R $ and $ H > 0 $, etc.

Since $ \tilde{\gamma} = \sqrt{ \exp(-2\tilde\eta_0) + H^2R^2} $, the bounds on $ \bar\lambda $, delineating the sub-, critical and super-critical regimes, depend on two parameters: $ \tilde\eta_0 $, which measures the gravitational C-energy inside the shell --- see equation~\eqref{eq:eta0Tilde_init_a} --- and $ HR $, i.e.\ the axial expansion rate measured in units of inverse circumference $ R^{-1} $. If one assumes a flat initial profile $ \partial_{\tilde t}\tilde{\alpha} = \text{const.} $ and $ \partial_{\tilde r}\tilde{\alpha} = 0 $ as we did in our numerics, $ \tilde{\eta}_{0i} $ and $ HR $ are related via~\eqref{eq:eta0Tilde_init}, and so the bounds~\eqref{eq:D_bounds} can be plotted in a $HR$-$\bar{\lambda}$-diagram. This is shown in Fig.~\ref{fig:regionPlot}, where the region below the solid purple (dark gray) line corresponds to $ D^+ $, the region in between the two solid lines is $ D^\uparrow $ (or $ D^\downarrow $), and everything above the solid orange (light gray) line is $ D^- $.
\section{Co-moving coordinates} \label{ap:cigar_coord}
It is straightforward to check that after introducing new coordinates $ (\bar t, \bar r) $ according to
\begin{align}
	\bar t &= L \ln \left( \frac{L \tilde t}{r_+^2} \sqrt{1 - x^2} \right) \;,\\ 
	\bar r &= \frac{r_+}{2} \left( 1 + \frac{1}{\sqrt{1 - x^2}}\right) \;,
\end{align}
with $ x \equiv \tilde r / \tilde t $, the scaling solution~\eqref{eq:scaling_sol} takes the form
\begin{multline}\label{eq:Greg_coord}
	\rd \tilde s^2 = \frac{\bar r^2}{L^2} \left( -\rd \bar t^{2} + \re^{2\bar t/ L}\rd z^{2} \right) +\\
	 \left( 1 - \frac{r_+}{\bar r} \right)^{-1} \rd \bar r^2 + 4 r_+^2\left(1-\frac{r_+}{\bar r} \right) \rd\phi^2 \,.
\end{multline}
Here, $L$ is an arbitrary length scale  which can be adjusted by rescaling (and shifting) $\bar t$. The constant $r_+$ denotes the position of the (regular) axis in the new coordinates, and the shell is now sitting at a constant coordinate position $ \bar r_0 $. The two parameters $r_+$ and $\bar r_0$, determining the range of $ \bar r $, are related to $R$ and $\tilde v$ via
\begin{subequations}
\begin{align}
	r_+ & = R \frac{\left(1 + \sqrt{1 - \tilde v^2} \right )}{2\tilde v} \,, \\
	\bar r_0 & = R \frac{\left( 1 + \sqrt{1 - \tilde v^2} \right)^2}{4 \tilde v \sqrt{1 - \tilde v^2}}\,. \label{eq:r0bar}
\end{align}
\end{subequations}

The metric~\eqref{eq:Greg_coord} is exactly the 4D version of the one discussed by Witten in 5D~\cite{Witten:1982} and by Gregory for general dimensionality~\cite{Gregory:1995qh}.  The benefit of these coordinates lies in their simplicity which allows to read off the geometrical content directly from the metric. The entire dynamics consists in a de Sitter-like expansion in axial direction, whereas the radial profile --- and hence in particular the interior area --- is completely static. Note that the scaling solution is only a true attractor for the interior spacetime, i.e.\ the red region in Fig.~\ref{fig:embed_supercrit} and~\ref{fig:embed_supercrit_reg}. However, as discussed in the main text, the exterior geometry also approaches the scaling solution with $ \tilde v \to 1 $ sufficiently far away form the shell. Hence, we can use~\eqref{eq:Greg_coord} also to picture the spacetime outside the string, if we neglect a small region close to it. Then the axis at $ r_+ $ corresponds to the south pole in Fig.~\ref{fig:embed_supercrit} and \ref{fig:embed_supercrit_reg}. The case with a conical singularity is simply obtained by appropriately adjusting the coefficient of $ \rd\phi^2 $ in~\eqref{eq:Greg_coord}. Moving away from the axis towards the string, $ \bar r \to \infty $ (because in the limit $ \tilde v \to 1 $ eq.~\eqref{eq:r0bar} implies $ \bar r_0 \to \infty $) and so the physical circumference approaches a constant value. The resulting embedding picture corresponds to a cigar-shaped geometry. In the interior, $ \tilde v $ is bounded by $ 3/5 $ and so the ratio $ \bar r_0 / r_+ $ is always smaller than $ 9/8 $. At this point, one has not yet reached the vertical part of the cigar, and so the interior embedding geometry always corresponds to a nearly flat cap. These results nicely agree with what was found in Fig.~\ref{fig:embed_supercrit} and \ref{fig:embed_supercrit_reg}.

\section{Validity of EFT}\label{ap:EFT}
General Relativity viewed as an effective field theory (EFT) is valid up to the Planck scale $M_{\rm P}$. Once the curvature scale exceeds this value, higher order operators become important and we can no longer trust its classical predictions. Generically, this happens once the 4D energy density becomes of order $M_{\rm P}^4$. For a cosmic string this depends on both the tension $\lambda$ and the regularization scale $R$. A solid way to derive the regime of validity of the EFT consists in considering the extrinsic curvature of the string. More specifically, we focus on the combination $\mathcal{K}:= [K_{\hphantom{c}c}^c]-[K_{\hphantom{0}0}^0] $ which is determined by the $(0,0)$ component of the matching equation \eqref{eq:israel}:
\begin{align}\label{eq:EFT}
\frac{\mathcal{K}}{M_{\rm P}}=\frac{\bar \lambda}{R M_{\rm P}}\;.
\end{align}
Once $\mathcal{K}$ exceeds $M_{\rm P}$, we expect the EFT to break down, or equivalently, Quantum Gravity effects to become important. The super-critical solutions we described are valid for $\bar \lambda \sim \mathcal{O}(1)$. Then \eqref{eq:EFT} implies that classical GR is applicable if and only if $R$ is much larger than the Planck length $ L_{\rm P} \equiv \Mp^{-1} $. Alternatively, this result becomes obvious when we naively estimate the 4D energy density as the ratio $\lambda/R^2$ and require it to be smaller than $M_{\rm P}^4$. In other words, even though the energy per string length needs to be Planckian in order to enter the super-critical regime, the energy per string volume is sub-Planckian if the string's thickness is much larger than the Planck length.
Also note that the marginally super-critical solutions only cover the range $ HR < 3/4 $, implying that the expansion energy $ \Mp^2H^2 $ is always smaller than $ \Mp^2/R^2 $ and hence also sub-Planckian if $ R \gg L_\mathrm{P} $.

\newpage
\bibliography{Supercrit_v5}

\end{document}